\documentclass[aps,pra,onecolumn,superscriptaddress,notitlepage]{revtex4-2}
\usepackage{tikz}
\usetikzlibrary{calc}
\usetikzlibrary{shapes.geometric, arrows}
\usepackage{eufrak}
\usepackage{enumitem}
\usepackage{mathtools}
\usepackage{graphicx}
\usetikzlibrary{positioning}
\usepackage{bm}
\usepackage{xcolor}
\usepackage[justification=justified,singlelinecheck=false]{caption}
\newcommand{\bematrix}{\left(\begin{matrix}}
\newcommand{\ematrix}{\end{matrix}\right)}

\usepackage{ulem} % only here for strikethrough-font
\usepackage[caption=false]{subfig}
\usepackage{dcolumn}
\usepackage{amsmath,amssymb}
\usepackage{bm}
\usepackage{bbm}
\usepackage{overpic}
\usepackage{latexsym}
\usepackage{color}
\usepackage[english]{babel}
\usepackage{latexsym}
\usepackage{psfrag,graphicx}
\usepackage{epsf}
\usepackage{amsmath}
\usepackage{amssymb}
\usepackage{amsfonts}
\usepackage{natbib}
\usepackage{multirow} 
\usepackage{appendix}
\usepackage{verbatim}
\usepackage{enumitem}
\usepackage{dsfont}
\usepackage{float}
\usepackage{tikz}

\definecolor{mygrey}{gray}{0.35}
\definecolor{myblue}{rgb}{0.2,0.2,0.8}
\definecolor{myzard}{cmyk}{0,0,0.05,0}
\definecolor{mywhite}{rgb}{1,1,1}
\definecolor{myred}{rgb}{0.9,0.1,0.}
\usepackage[colorlinks=true,citecolor=myblue,linkcolor=myblue,urlcolor=myblue]{hyperref}

\usepackage[makeroom]{cancel}

\newenvironment{proof-of}[1]{\medskip\noindent\textbf{Proof of {#1}.}}{\hfill$\blacksquare$\medskip}
\newcommand{\ket}[1]{\left\vert#1\right\rangle}

\newcommand{\braket}[2]{\ensuremath{\langle #1 | #2 \rangle}}

\usepackage{colortbl}
\usepackage{xcolor}

\usepackage{tcolorbox}

\definecolor{lightgray}{gray}{0.9}

\begin{document}

\title{Bell state measurements in quantum optics: a review of recent progress and open challenges}

\author{Luca Bianchi}
\affiliation{Department of Physics and Astronomy, University of Florence, 50019, Firenze, Italy}

\author{Carlo Marconi}
\affiliation{Istituto Nazionale di Ottica - Consiglio Nazionale delle Ricerche (INO-CNR), Largo Enrico Fermi 6, 50125 Firenze, Italy}

\author{Davide Bacco}
\email{davide.bacco@unifi.it}
\affiliation{Department of Physics and Astronomy, University of Florence, 50019, Firenze, Italy}

\begin{abstract}
\noindent Bell state measurements, which project bipartite states onto the maximally entangled Bell basis, are central to a wide range of quantum information processing tasks, including quantum teleportation, entanglement swapping, and fusion-gate quantum computation. In photonic platforms, where information is encoded in optical degrees of freedom, the realization of efficient Bell state measurements is particularly challenging, especially when constrained to linear optical elements. In this review, we provide a comprehensive examination of existing proposals for the implementation of Bell state measurements, highlighting their fundamental limitations and the strategies developed to overcome them. Moreover, we survey recent advances and discuss open challenges, with a particular focus on Bell state measurements for high-dimensional systems, an area of growing interest due to its relevance for quantum repeaters and scalable quantum networks. 

\end{abstract}

\date{\today}

\maketitle

\section{Introduction}
The first quantum revolution gave rise to technologies such as transistors, lasers and integrated circuits, profoundly reshaping modern society. Building on this foundation, the second quantum revolution aims to exploit uniquely quantum phenomena, such as entanglement, to unlock new possibilities in information processing and communication.
Indeed, the role of entanglement as a resource in quantum information is evident across a wide range of protocols, including quantum teleportation \cite{bennett1993teleporting,pirandola2015advances}, superdense coding \cite{bennett1992communication,mattle1996dense}, and quantum key distribution (QKD) \cite{bennett2014quantum,pirandola2020advances,xu2020secure}, among others. Like any other resource, entanglement can be quantified using appropriate measures \cite{plenio2014introduction}, which naturally lead to the notion of maximally entangled states, that provide the greatest utility for certain tasks. For multipartite systems, identifying such states is a challenging task, since different entanglement measures may not agree on which states are the most entangled \cite{dur2000three,acin2001classification}. However, a striking simplification arises in the bipartite scenario, where any entanglement measure can be expressed as a function of the Schmidt coefficients of a given state \cite{vidal2000entanglement}. This feature allows for a unique definition of maximally entangled states which, up to local unitary transformations, correspond to the so-called \textit{Bell states} \cite{nielsen2001quantum}.
For qubit systems, the four Bell states are defined as
\begin{equation}
\label{eq:bell_qubit}
    \ket{\Psi^{\pm}} = \frac{1}{\sqrt{2}}(\ket{00}_{L} \pm \ket{11}_{L})~, \qquad \ket{\Phi^{\pm}} = \frac{1}{\sqrt{2}}(\ket{01}_{L} \pm \ket{10}_{L})~,
\end{equation}
where $\ket{i}_{L}\in \mathbb{C}^{2}$ denotes a logical qubit state, with $i\in\{0,1\}$. Such states form an orthonormal basis for the Hilbert space $\mathcal{H} = \mathbb{C}^{2} \otimes \mathbb{C}^{2}$, usually called \textit{Bell basis}.
A Bell state measurement (BSM) \cite{bertlmann2002bell} is a joint measurement on a pair of qubits that projects their state onto the Bell basis. The ability to perform a complete and efficient BSM is a fundamental requirement in several quantum information protocols.
For instance, in \textit{quantum teleportation}, BSM is used to transfer an unknown quantum state from one party to another using shared entanglement and classical communication \cite{boschi1998experimental,pirandola2015advances}. In \textit{entanglement swapping} \cite{zukowski1993event}, BSM allows two previously unentangled particles to become entangled, thereby extending the reach of entanglement across quantum networks \cite{pan1998experimental}. This principle underlies the operation of quantum repeaters \cite{azuma2023quantum}, which are essential for long-distance quantum communication. Moreover, in the context of \textit{fusion-gate} quantum computation \cite{ralph2002linear, bartolucci2023fusion}, Bell measurements enable the construction of large-scale entangled states by merging smaller entangled building blocks. Finally, in the context of \textit{quantum key distribution}, BSMs lie at the core of protocols that allow for establishing a secret key even when the measurement devices cannot be trusted \cite{lo2012measurement}.
In this review, we provide a comprehensive overview of BSMs, exploring both their theoretical foundations and experimental realizations, while emphasizing their crucial role in advancing quantum technologies. Given the versatility of photonic platforms for encoding quantum states, this review focuses on quantum optical implementations of BSMs, examining the advantages and limitations of various approaches and highlighting ongoing advances aimed at improving their efficiency.

This review is structured as follows. In section \ref{sec:qubit}, we recall some basic aspects of qubit systems and their representation in quantum optics. Section \ref{sec:BSM-qubit} provides an overview of the main techniques for implementing BSMs in qubit systems, discussing the theoretical limitations on the success probability and detailing various strategies to overcome these constraints. In section \ref{sec:HD}, we shift our focus to qudit systems, discussing the most common photonic degrees of freedom used for their encoding. Section \ref{sec:BSM-HD} reviews existing techniques for performing efficient high-dimensional (HD) BSMs. Since the previous sections focus on discrete-variable (DV) encodings, section \ref{sec:CV} addresses the case of continuous-variable systems (CV). In section \ref{sec:app}, we explore key applications of BSMs in quantum information, with an emphasis on recent experimental advances. Finally, in section \ref{sec:conclusions}, we outline several directions for future research.

\section{Qubit systems}
\label{sec:qubit}
In the framework of quantum optics, photonic qubit states can be described using two different encodings. In the so-called \textit{single-rail} encoding (see e.g., \cite{furusawa2011quantum}), the state of a qubit is represented within a single optical mode, e.g., in its Fock basis $\{\ket{0},\ket{1}\}$, so that a generic qubit state takes the form $\ket{\psi} = \alpha \ket{0} + \beta \ket{1}$, where $\alpha, \beta \in \mathbb{C}$ satisfy the normalization constraint $|\alpha|^2 + |\beta|^2=1$. However, this encoding is not particularly convenient for quantum optics, since it would require nonlinear interactions to implement simple transformations on single qubits \cite{furusawa2011quantum}. 
A more suitable representation of the state of a photonic system is given by the \textit{dual-rail} encoding, where two orthogonal modes are used. In this sense, common choices are the polarization modes (e.g., vertical and horizontal) or the two spatial modes of a beam splitter. Hence, using this encoding, the logical qubit states take the form $\ket{0}_{L} = \ket{10}, \ket{1}_{L} = \ket{01}$, so that the Bell states of Eq. (\ref{eq:bell_qubit}) can be expressed as 
\begin{equation}
    \label{eq:bell_qubit_dual}
    \ket{\Psi^{\pm}} = \frac{1}{\sqrt{2}}(\ket{1010} \pm \ket{0101})~, \qquad \ket{\Phi^{\pm}} = \frac{1}{\sqrt{2}}(\ket{1001} \pm \ket{0110}~.
\end{equation}
In the following, unless explicitly stated otherwise, all photonic states will be expressed in dual-rail encoding. 
Finally, we recall that dual-rail encoded photonic states can be equivalently expressed in terms of a set of bosonic operators $\hat{a}_{k},\hat{a}^{\dagger}_{k}$ such that $[\hat{a}_{k},\hat{a}_{k'}]=[\hat{a}^{\dagger}_{k},\hat{a}^\dagger_{k'}]=0$ and $[\hat{a}_{k},\hat{a}^{\dagger}_{k'}] =\delta_{k,k'}$, where $\delta_{k,k'}$ is the Kronecker delta between two optical modes and $\hat{a}_{k}, \hat{a}^{\dagger}_{k}$ denote the operators annihilating and creating a photon on mode $k$, respectively. Hence, Eq.(\ref{eq:bell_qubit_dual}) can be cast as
\begin{equation}
    \label{eq:bell_qubit_op}
    \ket{\Psi^{\pm}} = \frac{1}{\sqrt{2}}(\hat{a}^{\dagger}_{1}\hat{a}^{\dagger}_{3} \pm \hat{a}^{\dagger}_{2}\hat{a}^{\dagger}_{4})\ket{\text{vac}}~, \qquad \ket{\Phi^{\pm}} = \frac{1}{\sqrt{2}}(\hat{a}^{\dagger}_{1}\hat{a}^{\dagger}_{4} \pm \hat{a}^{\dagger}_{2}\hat{a}^{\dagger}_{3})\ket{\text{vac}}~,
\end{equation}
where $\ket{\text{vac}}$ denotes the total photonic vacuum state.

\section{Bell state measurements for qubit systems}
\label{sec:BSM-qubit}
Due to the absence of photon-photon interactions at low energy scales, the implementation of entangling gates by means of linear optical devices, auxiliary states and post-selection can be realized only probabilistically. This approach, originally proposed by Knill, Laflamme and Milburn, is known as KLM quantum computation \cite{knill2001scheme}. In \cite{braunstein1995measurement,weinfurter1994experimental}, a different approach for the realization of a quanutm-optical BSM  was proposed and experimentally realized in \cite{bouwmeester1997experimental}, where it was used to implement a quantum teleportation protocol. Here, rather than relying on the construction of a conditional CNOT gate, the success of the protocol relies on the interference and indistinguishability of photons.
\begin{figure}[h]
    \centering
    \includegraphics[width=0.5\linewidth]{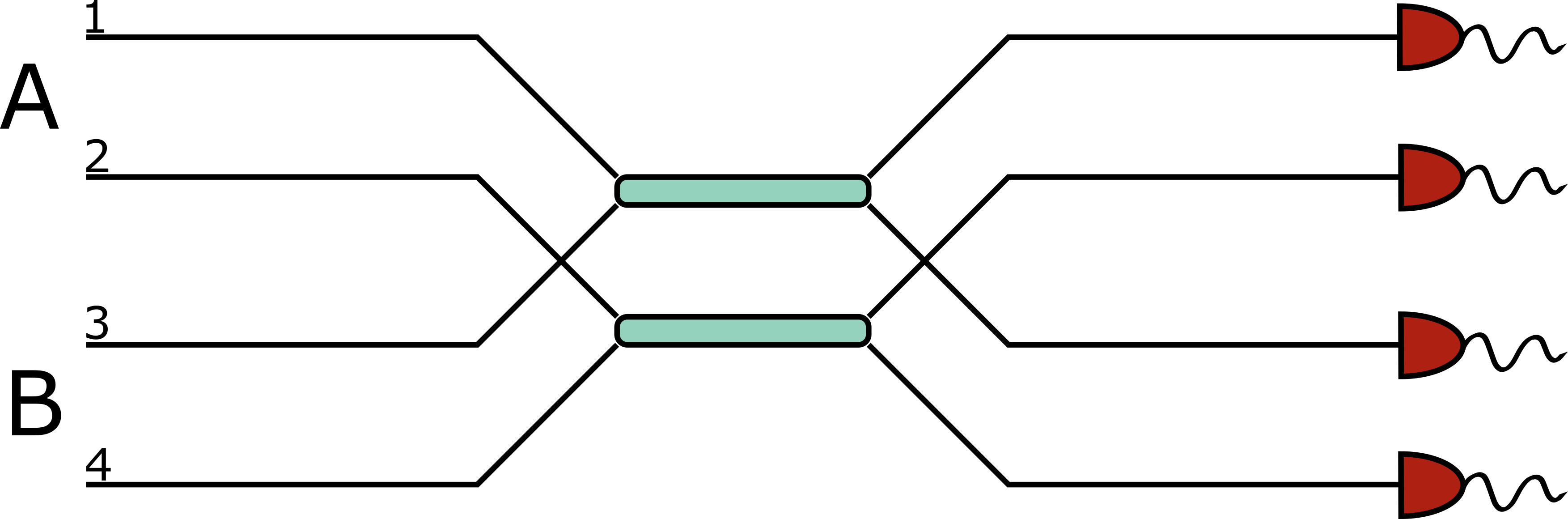}
    \caption{
    Simplest linear-optical BSM in discrete-variable quantum information. Two qubits, $A$ and $B$, are encoded in spatial modes: paths $1$, $2$ for qubit $A$, paths $3$, $4$ for qubit $B$. A global unitary transformation is implemented using two beam splitters coupling paths $1-3$ and $2-4$, respectively. Hong-Ou-Mandel interference enables entanglement between $A$ and $B$. Photodetectors record click patterns, of which only a subset uniquely identifies a given Bell state. Summing the probabilities of these unambiguous outcomes yields a success probability of $1/2$, which Ref.~\cite{lutkenhaus1999bell} showed to be the upper bound for a two-qubit system undergoing a generic linear interferometer with post-selection and photon detection.}
    \label{fig:BSMstandard}
\end{figure}
Typically, the experimental implementation of a measurement in the Bell basis exploits the Hong-Ou-Mandel (HOM) effect \cite{hong1987measurement}, by superimposing two initially uncorrelated photons on a $50/50$ beam splitter (see Fig.~\ref{fig:BSMstandard}). Then, single-photon detections are performed at the output of the beam splitters. A coincidence measurement between the four detectors signals a successful BSM, corresponding to the collapse of the quantum state into one of the Bell states of Eq.(\ref{eq:bell_qubit}). However, this simple scheme can distinguish only two of the four Bell states, thus resulting in an overall $50\%$ efficiency for this discrimination task.
For some time, it remained unclear whether the inherently probabilistic nature of this process was a consequence of the specific experimental implementations or, instead, was rooted in a deeper physical limitation. This ambiguity was resolved in \cite{lutkenhaus1999bell, vaidman1999methods}, where it was demonstrated that a complete BSM (that is, with $100\%$ success probability) cannot be achieved using only linear-optical components. This naturally raised the question of whether there exists a fundamental limit on the success probability. In \cite{calsamiglia2001maximum}, the authors showed that, in the case of a dual-rail encoding, the success probability of a BSM is theoretically bounded from above by $50\%$, thereby confirming the optimality of earlier experimental schemes. It is instructive to briefly recall the derivation of the above result. 
In what follows, we consider two photonic qubits encoded in the polarization degree of freedom. For an optical system with $m$ modes, a generic two-photon input state can be written as 
\begin{equation}
    \ket{\Psi_\mathrm{in}} = \sum_{i,j=1}^{m} N_{ij} \hat{a}_i^\dagger \hat{a}_j^\dagger \ket{0} 
    = \hat{\bm{a}}^\top N \hat{\bm{a}} \ket{0},
\end{equation}
for some $m \times m$ symmetric matrix $N$, where we have introduced the vector $\hat{\bm{a}} = (\hat{a}_1^\dagger,\dots, \hat{a}_m^\dagger)$.
Let us now observe that, since we are restricting to linear optical components, the action of any such device can be described in terms of a unitary matrix $U$, whose action on the input modes is given by
\begin{equation}
    \hat{c}_i^\dagger = \sum_{j=1}^{m} U_{ji}^* \hat{a}_j^\dagger~,
\end{equation}
where $\hat{c}, \hat{c}^{\dagger}$ denote the output modes after the transformation.
Equivalently, the input state $\ket{\Psi_{\rm in}}$ can be written as
\begin{equation}
    \ket{\Psi_\mathrm{in}} = \hat{\bm{c}}^\top M \hat{\bm{c}} \ket{0}~, 
\end{equation}
where $\hat{\bm{c}}= (\hat{c}^{\dagger}_{1},\dots, \hat{c}^{\dagger}_{m})$ and $M = U^\top N U$.
Let us now denote the Bell states as $\{\ket{\Psi^{(\mu)}}\}_{\mu=1}^4$, whose corresponding matrices, after the transformation $U$, are given by $M^{(\mu)} = U^\top N^{(\mu)} U$. Next, we introduce the $4 \times m$ matrix $U_{\rm tr}$ which is obtained by truncating $U$ so that its four rows correspond to the modes that encode the input photons, i.e., $U_{\mathrm{tr}} = (\bm{u}_1,\dots,\bm{u}_m)$~, where $\bm{u}_i=(U_{1i},U_{2i},U_{3i},U_{4i})$. Finally, introducing the $4\times4$ matrix $\Psi^{(\mu)}$ associated to the Bell state $\ket{\Psi^{(\mu)}}$, i.e., 
\begin{equation}
    \Psi^{(\mu)} =
\begin{pmatrix}
0 & 0 & \delta_{\mu1} + \delta_{\mu2} & \delta_{\mu3} + \delta_{\mu4} \\
0 & 0 & \delta_{\mu3} - \delta_{\mu4} & \delta_{\mu1} - \delta_{\mu2} \\
\delta_{\mu1} + \delta_{\mu2} & \delta_{\mu3} - \delta_{\mu4} & 0 & 0 \\
\delta_{\mu3} + \delta_{\mu4} & \delta_{\mu1} - \delta_{\mu2} & 0 & 0
\end{pmatrix}~,
\end{equation}
we can define $\bm{s}_i^{(\mu)} = \Psi^{(\mu)} \bm{u}_i$ as the vector representation of the output Bell state, that is, after applying the transformation $U$ on $\ket{\Psi^{(\mu)}}$. The probability of detecting two photons in a single output mode $\hat{c}_i$ for input $\ket{\Psi^{(\mu)}}$ is given by
\begin{equation}
    P_i^{(\mu)}(2) = \frac{1}{4} |\bm{u}_i^\top \Psi^{(\mu)} \bm{u}_i|^2~.
\end{equation}
By requiring that a two-photon detection event uniquely identifies a Bell state, it follows that $P_i^{(\mu)}(2) = 0$ for three of the four $\mu$.
After some algebra, we see that the only vectors $\bm{u}_i$ satisfying this condition are of the form $(a,b,0,0)$ or $(0,0,a,b)$. However, both cases actually lead to $P_i^{(2)}(\mu)=0$ for all $\mu$.
As a consequence, two-photon detection events never provide unambiguous discrimination.

Turning to single-photon detection, we are left with
\begin{equation}
    \ket{\Psi_i^{(\mu)}} = 2 \sum_{j\ne i=1}^{m} (M^{(\mu)})_{ij} ~\hat{c}_j^\dagger \ket{0} 
    = \frac{1}{\sqrt{2}} U_{\mathrm{tr}}^\top \Psi^{(\mu)} \bm{u}_i \cdot \hat{\bm{c}} \ket{0}~,
\end{equation}
where $\ket{\Psi_i^{(\mu)}}$ is the conditional state given by detecting one photon in mode $\hat{c}_i$ from the input state $\ket{\Psi^{(\mu)}}$.
We can check that the four vectors $\{\bm{s}_i^{(\mu)}\}$ are linearly dependent, i.e.,
\begin{equation}
    \det(\bm{s}_i^{(1)}, \bm{s}_i^{(2)}, \bm{s}_i^{(3)}, \bm{s}_i^{(4)}) = 0, 
\end{equation}
implying that
$\sum_{\mu=1}^4 b_\mu \bm{s}_i^{(\mu)} = 0 $
with at least two $b_\mu \neq 0$. As a consequence, at most two of the conditional states 
$\left\{\ket{\Psi_i^{(\mu)}}\right\}$ can be linearly independent, meaning
that, at most, only two Bell states can be unambiguously discriminated by detection in mode $i$, i.e., single clicks in different modes.
This implies that also the conditional states are linearly dependent, and the overlap between two of them is given by
\begin{equation}
    \braket{\Psi_i^{(\nu)}}{\Psi_i^{(\mu)}} = \frac{1}{2}\left[ \bm{s}_i^{(\nu) \, *}\bm{s}_i^{(\mu)} -(\bm{u}_i\cdot \bm{s}_i^{(\nu) \, })^*(\bm{u}_i \cdot\bm{s}_i^{(\mu)})\right]~,
\end{equation}
where the symbol $*$ denotes the complex conjugation. By setting $\nu = \mu$, we obtain the probability for a one-photon detection in mode $i$ 
\begin{equation}
    P_i^{(\mu)}(1) = \frac{1}{2}(|{\bm{u}_i}|^2 - |\bm{u}_i \cdot \bm{s}_i^{(\mu)}|^2).
\end{equation}

Let $i$ and $j$ denote the modes in which single-photon detections occur during a successful event, and choose $a$ and $b$ so that exactly two states can be discriminated.
For mode $i$, the probability that it contributes to a successful discrimination is bounded by
\begin{equation}
    P_{s,i} \le \frac{1}{4} ( P_i^{(a)}(1) + P_i^{(b)}(1) ) \le \frac{1}{4} |\bm{u}_i|^2,
\end{equation}
where the factor $1/4$ derives from assuming a uniform probability distribution for the input Bell states.
Each successful event contributes to two modes so, after correcting for double counting we have
\begin{equation}
    P_s \le \frac{1}{2} \sum_{i=1}^{m}  P_{s,i} 
         \le \frac{1}{8} \sum_{i=1}^{m} |\bm{u}_i|^2.
\end{equation}
Because $U_{\mathrm{tr}}$ is a $4\times m$ submatrix of a unitary, the orthogonality of its columns give
\begin{equation}
    \sum_i |\bm{u}_i|^2 = \sum_{i=1}^m \sum_{j=1}^4 |U_{ji}|^2 = 4.
\end{equation}
Therefore,
\begin{equation}
    P_s \le \frac{1}{2}.
\end{equation}

The above result is further strengthened in Ref. \cite{calsamiglia2002generalized}, which embodies the qubit case in the more general HD scenario (see Sec. \ref{sec:BSM-HD}).

Within this theoretical upper bound, many experimental and practical challenges remain open. One area of active experimental research concerns the design of protocols with improved HOM visibility \cite{hong1987measurement}. This feature is a key aspect of probabilistic BSMs, as it can affect the overall quality of the considered quantum protocol, see \cite{basso2019entanglement} for a greatly-detailed analysis and \cite{basso2021quantum, laneve2025quantum} for experimental proofs.
We briefly recall that, due to the HOM effect, two indistinguishable photons can only bunch together at the output of a beam splitter, where they are detected by the same detector. Distinguishable photons, instead, result in detections on any photodetector, possibly giving single or multiple clicks. 
As a consequence, when the rate of the coincidences at the detectors, $C_{\mathrm{click}}$, is zero, only one photodetector has clicked, indicating that the input photons were identical. 
This feature is related to what is known as HOM visibility, defined as the relative ratio between the maximum and the minimum rates of measured coincidence clicks, i.e.,
\begin{equation}
    V = \frac{C_{\mathrm{click}}^{\mathrm{(max)}}-C_{\mathrm{click}}^{\mathrm{(min)}}}{C_{\mathrm{click}}^{\mathrm{(max)}}}.
\end{equation}
Here, $V=1$ corresponds to perfect indistinguishability, while $V=0$ signals the absence of quantum interference and, therefore, perfect distinguishability. 
In recent years, advances in BSM protocols with a focus on improving HOM visibility have been reported in \cite{tomm2021bright,zhai2022quantum}, reaching a maximum value of $V \approx 98\%$ \cite{ding2025high}.

Another area of active experimental research concerns efforts to improve detector efficiency. Photodetectors are devices that convert incoming light into an electrical signal, enabling the detection and measurement of photons.
One can distinguish between single-photon detectors, which provide an on/off response to the presence of photons, and photon-number-resolving detectors, which are able to count the number of photons. Photon-number-resolving detectors can resolve up to several photons (see, for instance, \cite{kong2024large, cheng2023100, ding2025photon} for recent progress), but single-photon detectors are sufficient in the context of standard BSMs, as described in this section.
Of particular practical relevance are superconducting nanowire single-photon detectors \cite{you2020superconducting}. In these devices, an ultra-thin superconducting wire is cooled below its critical temperature and biased with a current close to its critical value. When a photon is absorbed, it briefly breaks superconductivity, creating a resistive segment that produces a measurable voltage pulse, thereby signaling the arrival of a single photon.
A major challenge in photodetection is achieving high fidelity between the detected and the initial state. This is due to two main causes: dark counts, i.e., false detection events that occur even when no photon arrives, usually caused by thermal fluctuations or electronic noise inside the detector; and photon losses, which occur when an incoming photon is not detected due to imperfect detector efficiency or because the photon is absorbed elsewhere.
Research on photodetection has become a very active topic in recent years (see, for example, \cite{venza2025research} and references therein).

While all the previously mentioned approaches are based on a setup similar to that shown in Fig.~\ref{fig:BSMstandard}, we conclude by presenting two alternative approaches where, although the total success probability remains bounded by $50\%$, different techniques are employed.
In \cite{asenbeck2024hybrid}, a hybrid scheme combining homodyne conditioning with single-photon detection was shown to improve the fidelity of quantum teleportation and entanglement-swapping protocols.
In \cite{van2006hybrid}, a protocol was proposed and later experimentally validated in \cite{van2006experimental}, where the distinguishability is distributed among three of the four Bell states.

\subsection{Boosting the success probability of qubit Bell state measurements}
Here, we analyze how the bound on the success probability derived in the previous section can be circumvented by exploiting nonlinearities, which we classify into three main categories: auxiliary photons, nonlinear interactions and hyper-entanglement. In the following, we examine the corresponding theoretical approaches, as well as their experimental demonstrations, where available, providing a detailed discussion of their advantages and main limitations.

\subsubsection{Auxiliary states}

The first theoretical proposal to use auxiliary photons in the context of BSM for qubit systems appeared in \cite{grice2011arbitrarily}. In that work, $N-1$ entangled photonic states $\ket{\Gamma_{j}}$, with $j=1,\dots,N-1$, were employed alongside a standard Bell measurement setup to identify an unknown Bell state $\ket{\zeta}$. The core of this approach is to introduce additional quantum correlations via auxiliary photons, enhancing the fundamental efficiency of the measurement scheme in the discrimination of all four Bell states. Each of the $N-1$ auxiliary photons is described by an entangled state of the form
\begin{equation}
\label{eq:ancilla_grice}
    \ket{\Gamma_j} = \frac{1}{\sqrt{2}}\left(\hat{a}^{\dagger}_{H, 2^{j}+1}\dots\hat{a}^{\dagger}_{H, 2^{j+1}} + \hat{a}^{\dagger}_{V, 2^{j}+1}\dots \hat{a}^{\dagger}_{V, 2^{j+1}}\right)\ket{\text{vac}}~,
\end{equation}
where $\hat{a}^{\dagger}_{X, k}$ is the bosonic creation operator for a photon in mode $k$ with polarization $X \in \{H,V\}$. 
Since each $\ket{\Gamma_j}$ contributes an even number of photons, all with identical polarization, the overall parity of the global input state $\ket{\zeta} \ket{\Gamma_{1}} \cdots \ket{\Gamma_{N-1}}$ remains conserved. As a consequence, the Bell states $\ket{\Psi^{\pm}}$ and $\ket{\Phi^{\pm}}$, which can already be distinguished without auxiliary photons from the parity of their detection patterns, can still be discriminated by counting the number of horizontally ($n_H$) and vertically ($n_V$) polarized photons. Specifically, odd values of $(n_{H}, n_{V})$ indicate the Bell states $\ket{\Psi^{\pm}}$, while even values correspond to the states $\ket{\Phi^{\pm}}$. Furthermore, the increase in success probability is related to the quantity $n_H - n_V$: certain click patterns, which are uniquely associated to $\ket{\Phi^{+}}$, yield even values of $n_H - n_V$, whereas others, uniquely identifying $\ket{\Phi^{-}}$, result in odd values. Perfect discrimination fails only in the degenerate case when $n_H-n_V = \pm 2^{N}$, which occurs with a total probability of $1/2^N$. Thus, the overall success probability scales as $P_s = 1 - 1/2^N$, demonstrating an exponential improvement with the number $N$ of auxiliary entangled photons. Hence, for $N=2$, a success probability of $75\%$ was derived, going beyond the fundamental limit imposed by standard linear optics.
A detailed numerical analysis of the success rate of the above protocol under realistic conditions has been conducted in \cite{wein2016efficiency}, where several experimental parameters, such as photon losses, detector dark counts, and detector efficiency, have been taken into account into the simulations.

One of the main experimental limitations of the method proposed in \cite{grice2011arbitrarily} is that, in order to reach near-unity success probabilities, it is necessary to prepare auxiliary photons in a highly entangled state of the form of Eq.(\ref{eq:ancilla_grice}). Although such states can be generated via spontaneous parametric down-conversion (SPDC), their preparation is highly nontrivial and inherently probabilistic, therefore limiting the overall effectiveness of the protocol. \\
One way to overcome this drawback was presented in \cite{ewert20143}, where it was demonstrated that a $75\%$ success probability can be reached using as a resource two single-photon identical states undergoing a balanced beam splitter transformation. Then, the following auxiliary state is obtained through HOM effect:
\begin{equation}
\label{eq:ancilla_vanloock}
    \ket{\Gamma} = \frac{\ket{20} + \ket{02}}{\sqrt{2}}~.
\end{equation}
Following a scheme similar to the one in \cite{grice2011arbitrarily}, the authors of \cite{ewert20143} discuss the case of an arbitrary number of auxiliary photons of the form
\begin{equation}
    \ket{\Gamma_j} = \frac{1}{\sqrt{2}2^{2j-2}}\left[\prod_{\substack{k = 2^j + 1 \\ k \; \text{odd}}}^{2^{j+1}}\left(a^{\dagger}_k\right)^2 + \prod_{\substack{k = 2^j + 1 \\ k \; \text{even}}}^{2^{j+1}}\left(a^{\dagger}_k\right)^2 \right]\ket{\text{vac}}~,
\end{equation}
which can similarly be constructed by repeated HOM effect on a network of beam splitters. 
The success probability in this case is given by $P_{s} = 1 - 1/2^{N+1}$, which approaches to unity when $N \rightarrow \infty$, thus allowing for an asymptotic deterministic BSM.
It should be noticed that the advantage of this proposal over the one presented in \cite{grice2011arbitrarily} lies in the $N=1$ case, where the $75\%$ limit is, in principle, easily achievable. However, as $N$ increases, the complexity of the auxiliary state proposed in \cite{ewert20143} grows exponentially.
Taking advantage of the $N=1$ case, successful experimental implementations of this scheme were recently reported in \cite{bayerbach2023bell} and \cite{hauser2025boosted}, achieving success probabilities of $P_s = 57.9\%$ and $P_s = 69.3\%$, respectively. The same scheme was also employed for quantum teleportation in \cite{d2025boosted}.

In \cite{olivo2018investigating}, the authors examined the optimality of several auxiliary-based BSM schemes and proved an analytical upper bound on the success probability for polarization-preserving interferometers. Moreover, they revealed how this bound relates the complexity of the ancilla to the achievable performance. Through numerical optimization techniques, they also provided evidence that the approaches presented in \cite{grice2011arbitrarily} and \cite{ewert20143} saturate this bound and are indeed optimal within a particular set of experimental constraints. 
Finally, we conclude this paragraph by presenting an alternative model, based on fusion gates \cite{browne2005resource}, to implement a qubit BSM starting from HD systems \cite{yamazaki2025linear}. Here, the concept of \textit{pairwise fusion gate} is introduced as a tool to project HD states of local dimension $d$, onto two-dimensional Bell states. In particular, the resulting success probability is given by $P_s = 1-1/d$ without the use of auxiliary photons, and can be improved to $P_s =1 - 1/d^{k+1}$ when employing $N = 2(2^k-1)$ auxiliary states. Notice that such probability is independent of the particular choice of the auxiliary states, the chosen linear optical scheme and the encoding scheme, thus highlighting the versatility of this protocol.

It should be noted, however, that while theoretically well developed, auxiliary-photons protocols typically rely on experimentally demanding highly-entangled states. Furthermore, as we will see in \ref{sec:BSM-HD}, the above techniques do not show a good scaling with the dimension of the Hilbert space where the information is encoded.

\subsubsection{Nonlinear optics}
Nonlinear optical interactions offer another promising route to enhance the success probability of BSM schemes allowing them, in some cases, to be theoretically deterministic.
Among nonlinearities, squeezing stands out as one of the most versatile due to its wide applicability and experimental feasibility.
The first proposal to exploit squeezing for improving the success probability of a BSM was presented in \cite{zaidi2013beating}. There it was shown that, similarly to the case of auxiliary photons, squeezing increases the number of interference terms among initially uncorrelated photons. While this approach does not allow for unambiguous discrimination, as in the case of the previous protocol based on Kerr nonlinearity, it nonetheless provides a boost in the success probability for BSM, both in the case of single- and dual-rail encoding. 

In particular, considering a dual-rail polarization encoding, the setup involves an input Bell state impinging on a $50:50$ beam splitter, followed by two polarizing beam splitters. Single-mode squeezers, described by the unitary operator $\hat{S}(z) = \exp{\left\{ \left(z^*\hat{a}^2 - z\hat{a}^{\dagger \; 2} \right)/2\right\}}$ and parametrized by the (complex) squeezing parameter $z$, are then applied to each mode before detection with photon-number-resolving detectors (PNRDs). For the single-rail case, the scheme consists of a beam splitter, two single-mode squeezers, and a second beam splitter before the PNRDs.
In both cases, the improvement in the success probability is a function of the squeezing parameter $\zeta$, reaching a maximum $64.3\%$ at $\zeta \simeq 5.719 \; \text{dB}$ for dual-rail encoding, and $62.5\%$ at $\zeta \simeq 6.270 \; \text{dB}$ for single-rail. The increase in success probability is ensured by the parity-preserving nature of the squeezing interaction, which does not compromise the success probability achievable with linear optics. In fact, also when restricting to the simplest scheme of Fig.~\ref{fig:BSMstandard}, the unambiguous Bell states have a different parity from the ambiguous ones, implying that the inclusion of squeezing cannot perform worse than in this former case.
Furthermore, another advantage of this approach is that it eliminates the need for auxiliary states, thus avoiding the preparation of complex entangled states, whose generation remains a significant practical challenge.
A detailed numerical analysis of this scheme, including realistic considerations such as detector noise, limited resolution, and phase instability, was conducted in \cite{kilmer2019boosting}, where it was shown that the previously reported success probability of $64.3\%$ drops to a maximum attainable value of $59.6\%$ at $\zeta=0.5774$.

One of the first models for a deterministic BSM was presented in \cite{paris2000optical}, where a nonlinear interferometric approach for a polarization-encoded BSM was proposed. In this scheme, after impinging on a polarizing beam splitter, an unknown Bell state is sent through an interferometer where, in one arm, the polarization is rotated, while in the other, a 
quantum non-demolition measurement of the photon number is performed. Such measurement is based on the \textit{cross-Kerr effect}, a third-order nonlinear process described by an interaction of the form
\begin{equation}
\label{ham_kerr}
    \hat{H}_{\mathrm{Kerr}} \propto
            \hat{a}^{\dagger}\hat{a}~
            \hat{b}^{\dagger}\hat{b}~,
\end{equation}
where $\hat{a}, \hat{a}^{\dagger}$ refer to the polarization modes of the photon state, while $\hat{b},\hat{b}^{\dagger}$ to the cavity modes. After the interferometer, the photons are then recombined on a beam splitter and their coincidences are measured through single-photon photodetectors. In this way, one obtains a complete BSM, whose success is limited only by the efficiency of the nonlinear process. Subsequently, in \cite{vitali2000complete}, a simplification of the above scheme is presented, enhancing its feasibility using ultraslow light
propagation to achieve the necessary cross-Kerr nonlinearity.
On a similar note, in \cite{barrett2005symmetry} an alternative protocol is described, based on weaker cross-Kerr effects and thus requiring less amount of nonlinearity. However, in \cite{shapiro2006single}, it was shown that the Hamiltonian of Eq.(\ref{ham_kerr}) introduces decoherence effects that prevent the deterministic realization of high-fidelity entangling gates based on cross-Kerr effects.
Cross-Kerr effects have been also proposed to model probabilistic nonlinear optical gates, that might be used as fundamental building blocks for quantum computation (see, e.g., \cite{lin2009quantum, wang2012photonic}).

In \cite{kim2001quantum}, an alternative nonlinear approach was proposed and experimentally validated, where a BSM was implemented as part of a quantum teleportation protocol, relying on a nonlinear process known as sum-frequency generation (SFG). In SFG, two photons of different frequencies interact within a nonlinear medium to produce a single photon whose frequency equals the sum of the input ones. This process is described by a Hamiltonian of the form
\begin{equation}
    \hat{H} = ig\left( \hat a_1\hat a_2 \hat{b}^\dagger - \mathrm{h.c.} \right)
\end{equation}
where $g$ is the coupling strength. Remarkably, similarly to the case of protocols based on cross-Kerr nonlinearity, in \cite{kim2001quantum} there is no fundamental theoretical upper bound on the success probability of this implementation, and the only limitations are due to the efficiency of the SFG process itself. 
A similar technique was also employed in \cite{nloswap}, where it was experimentally applied to an entanglement swapping protocol.
Some notable drawbacks should be kept in mind while using nonlinear optics, the main one being the fact that this kind of interactions in optical systems often introduce additional noise, which can degrade the fidelity of the teleported state \cite{cesar2009extra}. Also, while the aforementioned methods are well-suited for polarization-encoded qubits, their extensions to other types of encoding is non-trivial, mainly due to the dependency on the power of the initial pump field. However, this may inspire future research aimed at adapting BSM protocols to different encodings, allowing each degree of freedom to be exploited to its optimal advantage.

We conclude this section by presenting another instance of nonlinear effects arising from the so-called \textit{light–matter} interactions an area that is attracting significant interest, particularly from the experimental community. So far, atomic systems have been treated using an effective description, where their influence on the photonic system is incorporated through coupling constants associated with specific nonlinear effects. However, a more accurate approach involves treating light and matter on equal footing, modeling the matter component as a collection of two-level quantum systems. In this fully quantum picture, a common theoretical framework is given by the quantum Rabi model \cite{xie2017quantum}, and more specifically, by the Jaynes–Cummings model \cite{greentree2013fifty}, which stands at the core of cavity quantum electrodynamics (cQED) \cite{walther2006cavity}. Based on this,
several techniques for deterministic BSMs have been proposed \cite{lloyd2001long} and experimentally realized \cite{specht2011single}, supporting the construction of scalable quantum photonic platforms \cite{duan2004scalable, reiserer2015cavity}. Similarly, \cite{zhou2015complete} introduces the use of \textit{Faraday rotations}, generalizing earlier proposals from physical to logical qubits, a scenario of great interest for error correction and long-distance quantum communication. Recently, experimental demonstrations of cQED–based BSMs have been reported in \cite{nolleke2013efficient,welte2021nondestructive, kamimaki2023deterministic}. We conclude this section by highlighting \cite{hofer2013time} where, in the context of light-matter interaction, an homodyne-like setup is presented for both continuous and discrete variables, potentially helping to overcome the probabilistic nature of all-photonic BSM for qubits (see Sec. \ref{sec:CV} for more details). This so-called \textit{time-continuous Bell measurement} ideally overcomes the $50\%$ probabilistic limit, and tolerates in principle photon losses up to $50\%$, making use of a local oscillator with a feedback mechanism.

\subsubsection{Hyper-entanglement}

Another strategy to deal with BSM is to embed qubit systems within a larger Hilbert space. This idea stands at the core of \textit{hyper-entanglement} \cite{kwiat1997hyper}, where a composite system is simultaneously entangled with respect to multiple degrees of freedom, thereby enabling an increased channel capacity for quantum communication protocols. The general form of an hyper-entangled state corresponds to the tensor product of entangled states in different degrees of freedom, e.g., 
\begin{equation}
    \ket{\mathrm{\Psi}} = \frac{1}{\sqrt{2}}\left(\ket{H}_1\ket{V}_2 \pm \ket{V}_1\ket{H}_2\right)\otimes\frac{1}{\sqrt{2}}\left(\ket{a}_1\ket{b}_2 + \ket{b}_1\ket{a}_2\right),
\end{equation}
where $H,V$ denote polarizations and $a,b$ refer to spatial modes. Although the global state factorizes across different degrees of freedom, it remains entangled within each subspace of the total Hilbert space. 
From a physical perspective, SPDC naturally generates hyper-entangled states. In particular, energy conservation in the nonlinear interaction implies energy-time entanglement in the down-converted photons. Momentum conservation (i.e., phase-matching conditions) further requires correlated emission directions, thus leading to spatial entanglement. 
Since energy-time correlations are inherently present in SPDC photons, additional degrees of freedom can be exploited by appropriately selecting spatial modes or engineering the polarization through suitable phase-matching configurations. In this way, entanglement can be generated across multiple degrees of freedom, giving rise to hyper-entangled states. Remarkably, such degrees of freedom are, in principle, independently addressable, so that any measurement on one of them can be performed without disturbing the others. This allows the entanglement in one degree of freedom to be used as an ancilla for the analysis of entanglement in another.

Based on this observation, \cite{kwiat1998embedded} introduced two proposals in which a two-qubit system is entangled with respect to polarization and an additional degree of freedom, namely time-energy or momentum. In this way, a scheme for a complete BSM was presented, based on hyper-entanglement and photon-number-resolving detectors. In \cite{wei2007hyperentangled} it was shown that, in the case of $n=2$ degrees of freedom, it is possible to distinguish only $7$ out of the $16$ hyper-entangled Bell states, when restricting to linear optics. Moreover, in the same work, the maximal number of mutually distinguishable such states was conjectured to be upper bounded by $2^{n+1}-1$ when only one copy of the Bell state is available. Further, complete distinguishability of all $4^n$ hyper-entangled Bell states is possible when two copies of the state are available. A general proof of this bound was found in \cite{pisenti2011distinguishability}. 
A different approach was presented in Ref.~\cite{walborn2003hyperentanglement}. Although this scheme does not theoretically achieve a fully complete BSM, it enhances the manipulation of independent degrees of freedom in such a way that PNR detectors are no longer required; instead, an exchange of classical communication between the two parties allows the BSM to be implemented nonlocally.
The main drawback of this method is that it can be applied only to down-converted photons resulting from an SPDC process.
In \cite{schuck2006complete}, the first deterministic BSM based on hyper-entanglement in polarization and energy-time was proposed. Notably, their setup required only linear optical components, without the need for auxiliary photons. Expanding on this, in \cite{barbieri2007complete}, an alternative protocol for deterministic BSM based on polarization-momentum hyper-entanglement was presented and experimentally realized. In this work, two SPDC photons in a type-I BBO crystal are entangled with respect to polarization and momentum. The momentum entanglement serves as an ancilla to discriminate polarization-encoded Bell states. The success of the protocol crucially depends on the preservation of the auxiliary momentum entanglement. With a combination of beam splitters, wave plates and polarizing optics, the two-photon polarization Bell states are mapped onto single-photon Bell states in the enlarged Hilbert space. The spatial modes are selected using a four-hole screen that defines the momentum qubit structure. An average fidelity of $F = 0.89$ is reported, with the main imperfections due to the reduced purity and coherence in the momentum degree of freedom. A further experimental proof of a hyper-entangled BSM was reported in \cite{wang2015quantum}, in the context of quantum teleportation.
Further works were devoted to explore how to distinguish hyper-entangled Bell states, i.e., tensor products of Bell states in two different degrees of freedom. 
Other works, such as \cite{sheng2010complete, wang2016complete} theoretically investigated the use of nonlinear effects within the problem of complete discrimination of hyper-entangled Bell states.
While a complete survey of other applications of hyper-entangled states in quantum information processing is out of the scope of this review, we refer the interested reader to \cite{deng2017quantum}.

\section{High-dimensional systems}
\label{sec:HD}
Qudit systems, whose corresponding Hilbert space has dimension $d$, offer several advantages over their qubit counterparts. One immediate benefit of encoding information in HD states is the potential reduction in the size and complexity of quantum circuits \cite{gedik2015computational,karacsony2024efficient}.
Specifically, a $D$-dimensional system requires $N_{\text{qubit}}$ qubits such that $2^{N_{\text{qubit}}} = D$, whereas only $N_{\text{qudit}}$ qudits are needed, with $d^{N_{\text{qudit}}} = D$. 
As a consequence, for large values of the local dimension $d$, fewer qudits are needed as compared to the qubit case, and the information storage capacity scales as $k = \log_2 d$.
Moreover, fewer transformations and controls are needed to manipulate the same amount of information \cite{wang2020qudits}.    

Operational lifetimes of qudits are influenced by noise and decoherence. Since each quantum gate requires a finite amount of time to implement, using fewer gates becomes crucial when operating within limited coherence windows. Moreover, it has been shown that any operation can be approximated by a unitary operator in $SU(d^{N_{\text{qudit}}})$, with an accuracy that scales as $1/d^2$ \cite{luo2014geometry}. This further reduces the number of gates required, enhancing overall coherence times.

From an experimental point of view, qudits have demonstrated superior information transmission rates as compared to qubits \cite{bacco2017space, bacco2019high}. They also exhibit enhanced resilience to noise, which is particularly beneficial in quantum key distribution (QKD) protocols \cite{sheridan2010security}.  In this context, robustness is often quantified by the quantum bit error rate (QBER), defined as the ratio of erroneous to total received bits \cite{nielsen2001quantum}. While a low QBER indicates reliable transmission, other metrics such as channel fidelity, gate error rates, and error-correction capabilities must also be considered. Notably, qudit systems exhibit higher error tolerance thresholds, allowing them to outperform qubits under equivalent QBER conditions \cite{bechmann2000quantum}. Furthermore, increasing the local dimensionality enables higher secure key generation rates within a fixed spatial region, though often at the expense of transmission range \cite{cerf2002security}.

Qudits also provide notable advantages for entanglement distribution \cite{ekert1991quantum}. Using fewer HD systems instead of many low-dimensional ones has been shown to mitigate the effects of noise \cite{liu2009decay}. Additionally, qudit dimension affects the fidelity of quantum cloning processes. For example, symmetric cloning of qudits achieves a fidelity of $1/2 + 1/(d+1)$, indicating that higher dimensions make it more difficult for adversaries to replicate quantum information accurately \cite{navez2003cloning, bruss1999optimal}. From a practical standpoint, photons represent the best candidates to implement HD quantum states. At low energies, photons can be regarded as non-interacting with electromagnetic fields, and their weak coupling with the environment leads to long coherence times, making them excellent for qudit transport in free space or optical fibers \cite{o2009photonic}. Moreover, HD quantum states have demonstrated their potential in co-propagating alongside classical optical communication channels, which could facilitate the integration of quantum devices into existing telecommunications networks \cite{wu2025integration}.

\begin{figure}
    \centering
    \includegraphics[width=\linewidth]{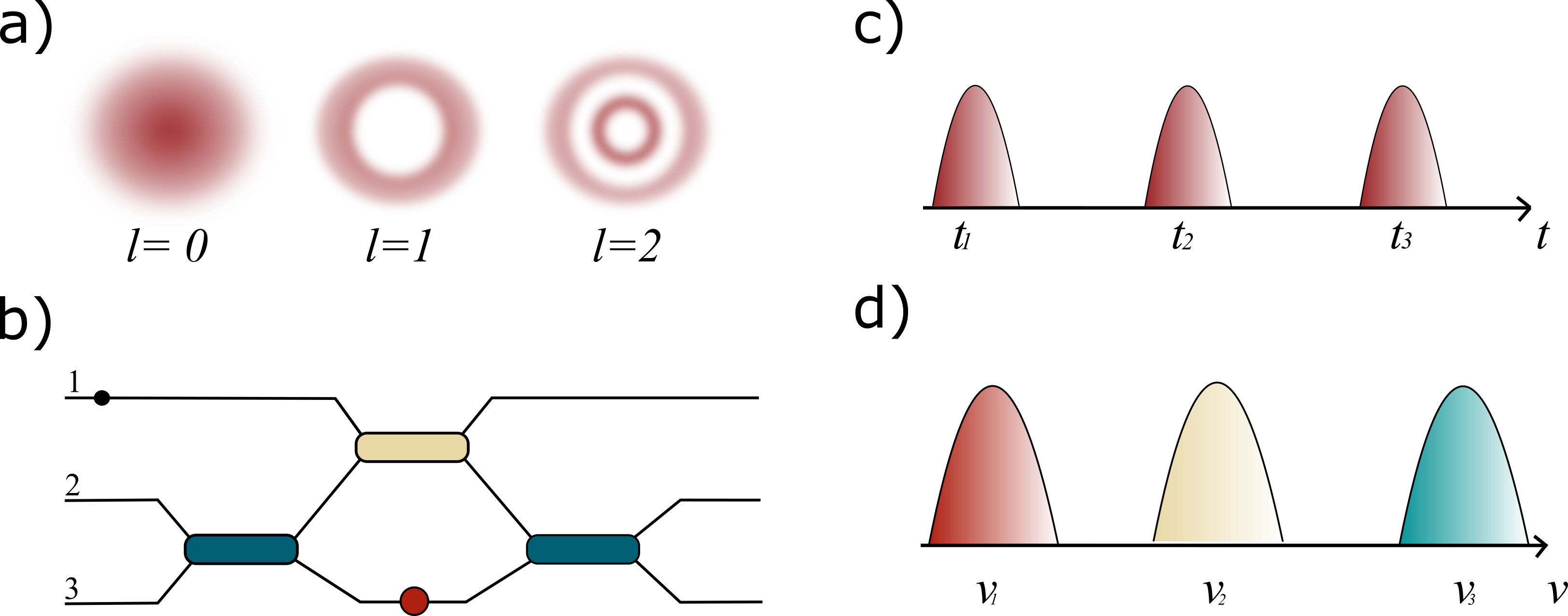}
    \caption{Pictorial representation of several degrees of freedom, commonly used to encode qudit systems. Here, we illustrate them for a qutrit, i.e., a three-dimensional system. a) Orbital angular momentum encoding: information is stored on angular momentum modes of a laser beam, and many families of modes may be employed. Changes in angular momentum can be applied through plates and modulators. b) Path encoding: photons can occupy one among many different paths, for instances different cores in an optical fiber. Photons can then be superimposed by means of beam splitters and phase shifters. Here, we report a quantum Fourier transform where balanced beam splitters (blue rectangles) are combined with another splitter with transmittivity $T = \sqrt{2/3}$ (yellow rectangle) and a phase shifter with shift $\phi = \pi/2$ (red circle). c) Time-bin encoding: photons may take different time to be detected. These different arrival times describe a basis in an Hilbert space, whose states might be manipulated with unitary transformations realized through delay-lines displaced in an optical circuit. d) Frequency encoding: different frequencies (displayed here through different colors) define different logical states. Operations may be taken by optical modulators.}
    \label{fig:encodings}
\end{figure}

Several photonic platforms support HD state encoding (see Fig.~\ref{fig:encodings}). Orbital angular momentum (OAM) encoding exploits the helical phase structure of light, where each quantized OAM mode, characterized by a topological charge $l \in \mathbb{N}_{0}$, forms a basis state in the qudit Hilbert space. Different modes may be employed, including Laguerre-Gauss \cite{mair2001entanglement}, Hermite-Gauss \cite{sephton2023quantum} and Bessel-Gauss \cite{mclaren2012entangled}. This method offers an unbounded state space and can be realized using devices like spiral phase plates or q-plates, making it a promising candidate for free-space communication and HD-QKD \cite{bouchard2018quantum}. However, OAM states are susceptible to atmospheric turbulence and experience significant mode dispersion in optical fibers, making it necessary the use adaptive optics to extend their transmission range \cite{ren2014adaptive} (for a review on OAM see \cite{erhard2018twisted,erhard2020advances, Mirhosseini2015}).
Path encoding is another widely adopted technique, wherein different spatial paths represent distinct qudit basis states. This encoding has found particular utility in integrated-photonics experiments, e.g., boson sampling \cite{spagnolo2014experimental}. Common implementations include integrated waveguides \cite{yang2025programmable}, interferometric networks \cite{khabiboulline2019optical}, and multicore fibers \cite{da2021path}. This approach enables high-fidelity state manipulation and is particularly well-suited to chip-scale integration, as demonstrated in universal photonic processors \cite{carolan2015universal}. However, implementing HD transformations requires increasingly complex optical networks \cite{ma2025preparation}, limiting scalability. 
Time-bin encoding \cite{brecht2015photon}, based on the arrival time of the photon, is implemented using interferometers with defined path-length delays. It offers excellent robustness against decoherence and has been successfully employed in long-distance fiber-based quantum communication \cite{vagniluca2020efficient}. However, while time-bin qudits can travel through standard optical fibers with minimal loss, the approach demands precise timing resolution and offers limited gate speed due to inherent temporal constraints \cite{yu2025quantum}.
An alternative option is represented by frequency encoding, where basis states are associated with discrete spectral components of photons \cite{Lukens2017, chang2023towards, chang2024time, chang2025recent}. This method benefits from compatibility with existing telecom infrastructure and allows efficient generation of entangled qudit states using standard optical tools \cite{jin2016simple, chang2021648}.
Each encoding approach offers specific strengths tailored to different applications: OAM allows for generating relatively easily qudit states of larger local dimension, path encoding suits integrated quantum processors, time-bin encoding is ideal for long-distance fiber-based systems, and frequency encoding aligns well with classical networks. However, each also has inherent limitations. To address these, hybrid encoding schemes have been proposed, where an increased versatility is reached at the cost of greater experimental complexity.
For instance, combining OAM and path encoding demonstrated enhanced state manipulation capabilities \cite{jo2019efficient}, while time-frequency entangled qudits have shown promise for robust quantum communication networks \cite{cheng2023high}. 

\section{High-dimensional Bell state measurements}
\label{sec:BSM-HD}
When the Hilbert space of the quantum system is enlarged to higher dimensions, the notion of Bell states should be properly generalized. To this end, two-qudit Bell states have been defined as \cite{bennett1993teleporting, sych2009complete}
    \begin{equation}
        \ket{\Psi_{lm}} = \frac{1}{\sqrt{d}}\sum_{k=0}^{d-1}\omega_d^{lk}\ket{k}_L\ket{(k\oplus m)}_L~,
        \label{eq:bell_qudit}
    \end{equation} 
where $l,m =0, \dots, d-1$, $\omega_{d}=\exp(2\pi i/d)$ is the $d$-th root of the unity and $\oplus$ denotes the sum modulo $d$. Different from the case of qubit systems, where all Bell states are either symmetric or anti-symmetric with respect to the exchange of the subsystems, in the HD scenario, there exist states that do not have a definite symmetry. More importantly, \cite{calsamiglia2002generalized} demonstrated that, under the constraints of linear optics, it is fundamentally impossible to perform an HD BSM even when ambiguous (i.e., probabilistic or non-deterministic) discrimination is allowed. The key steps in this proof consist of deriving an exact, closed-form expression for the POVM elements associated with a linear-optical transformation acting on a two-boson state. From this, it follows that such POVMs can have at most Schmidt rank $r=2$, whereas HD Bell states have Schmidt rank $r=d$. This establishes a direct link between the entangling power of a POVM and the degeneracy of click patterns associated with different input Bell states. Similarly to the qubit case, this no-go theorem can be circumvented by introducing nonlinear optical interactions or auxiliary photons. However, unlike in the qubit scenario — where such techniques are employed to enhance the success probability — they are strictly required in the HD case to achieve any nonzero success probability at all. This result was further strengthened in \cite{carollo2001role}, where it was showed that the inclusion of auxiliary photons can only enhance the success probability of ambiguous discrimination, although perfect discrimination remains impossible. Despite this severe limitation, qudit systems keep raising a considerable amount of interest for their benefits, especially in quantum communication \cite{bacco2019high}. Therefore, it is not surprising that many of the techniques presented in Sec.~\ref{sec:BSM-qubit} have been generalized to the HD setting, making the quest for qudit BSMs a rapidly evolving and active area of research. In the following, we discuss in detail several approaches for implementing HD BSMs.

\subsection{Boosting the success probability for high-dimensional Bell state measurements}

In the following subsections, we review the main theoretical results for HD BSM protocols that have been proposed and experimentally demonstrated so far. These protocols can be either probabilistic, resulting in incomplete BSMs, or deterministic, with their efficiency depending on the performance of the underlying quantum process. For probabilistic protocols, the corresponding success probabilities are compared in Table~ \ref{tab:comp}.

\begin{table}[h]
        \centering
        \begin{tabular}{|c|c|c|c|c|}

            \hline\hline

            dim & $P_s^{\mathrm{(l.o.)}}$ & $P_s^{\mathrm{(aux)}}$ & $P_s^{\mathrm{(sq)}}$ \\

            \hline
            \multirow{3}{*}{$d=2$} 
        &  & $3/4$ \cite{grice2011arbitrarily} &  \\ \cline{3-3}
        & $1/2$ & $3/4$  \cite{ewert20143}& $\approx 0.57$ \cite{kilmer2019boosting}\\
        
         \hline
         
            \multirow{3}{*}{$d=3$} 
            &  &  $1/81$ \cite{luo2019quantum} &\\  
            \cline{3-3}
            & $0$ & $1/6$ \cite{ustun2025fusion}& $\approx 0.037$ \cite{bianchi2025predetection}\\

        \hline
         
            \multirow{3}{*}{$d=4$} 
            &  &  $1/1024$ \cite{luo2019quantum} &  \\  
            \cline{3-3}
            & $0$ & $1/8$ \cite{bharos2024efficient} &$\approx 0.021$ \cite{bianchi2025predetection}\\

            \hline
         
            \multirow{3}{*}{$d=5$} 
            &  &  $9.2 \times 10^{-5}$ \cite{luo2019quantum} &\\  
            \cline{3-3}
            & $0$ & $1/15$ \cite{ustun2025fusion}& $\approx 0.0134$ \cite{bianchi2025predetection}\\
        
            \hline  \hline
        \end{tabular}
         \caption{
           Comparison of theoretical success probabilities for qubits and qudits in different protocols. The success probability of protocols restricted to linear optical devices, $P_s^{\mathrm{(l.o.)}}$, has a maximum value of $1/2$ for qubits and $0$ for qudits \cite{calsamiglia2002generalized}. In the second column, the success probabilities for auxiliary states methods, $P_s^{\mathrm{(aux)}}$, are listed. For qubit systems with a single ancilla this limit may be surpassed, reaching a theoretical value of $3/4$.
           For qudits, results from  different methods are reported. The simulated success probability for squeezing-based approaches, $P_s^{\mathrm{(sq)}}$ is reported in the last column for the case of a photon-number resolution able to resolve up to five photons.}
        \label{tab:comp}
\end{table}

\subsubsection{Auxiliary states}
The generalization to qudit states of auxiliary-photon-based methods was first explored in \cite{duvsek2001discrimination}, where the KLM protocol is systematically applied to construct a non-deterministic—i.e., conditional—SWAP (CSWP) gate. This two-photon gate is a fundamental building block for performing Bell measurements in higher dimensions, but its implementation requires a growing number of modes and entangled auxiliary photons as the system dimension $d$ increases. The overall success probability of the scheme is directly tied to the success rate of the non-deterministic operation and is lower bounded by
\begin{equation}
    P_s \geq \left(\frac{\mathcal{M}}{\mathcal{M}+1}\right)^{2(d-1)^2}~,
\end{equation}
where $d$ is the dimension of the qudit system and $\mathcal{M}$ denotes a parameter defined by the number of auxiliary photons and modes used in the setup, corresponding to $2\mathcal{M}$ auxiliary photons and $4\mathcal{M}$ auxiliary modes, respectively.

Given the nondeterministic nature of the KLM protocol and the low probability of generating highly entangled auxiliary states, the pursuit of near-deterministic Bell measurements was largely set aside in favor of more experimentally feasible solutions. In \cite{luo2019quantum}, a scheme was proposed achieving a success probability scaling with inverse powers of the dimension $d$, using only $d-2$ auxiliary single photons. In this approach, the input uncorrelated particles and auxiliary photons are processed through a $d$-dimensional quantum Fourier transform, which acts as a multiport beam splitter \cite{zukowski1997realizable}, and is supplemented by a non-unitary transformation defined on a $d+1$-dimensional space. By employing post-selection, the scheme enables unambiguous discrimination of a single Bell state out of the $d^2$ possibilities. Moreover, it was shown that, within a teleportation protocol, additional operations on Bob’s side can further enhance the overall success probability. The method was analyzed theoretically for $d=3,4$ and experimentally implemented in the qutrit case. It is worth noting that the general scaling law of this scheme is still unknown, as it was simulated only for $d=3,4$ in the original work and for $d=5$ in \cite{ustun2025fusion}. Moreover, the success of the protocol crucially depends on the indistinguishability of the photons and on the quality of the single-photon sources. A practical implementation of the above protocol was simulated in \cite{bacco2021proposal}.

Despite the experimental challenges associated with this scheme — primarily due to the need for HD interferometric setups and to the use of single-photons as auxiliary states — it laid the foundation for the development of more advanced auxiliary-based strategies. In particular, \cite{bharos2024efficient} introduced a significant theoretical improvement in HD BSM by modeling an entangled-state analyzer specifically tailored for Hilbert spaces with even dimension. In this approach, the auxiliary resource is an entangled state of the form:
\begin{equation}
\label{eq:ancilla_bharos}
    \ket{\text{aux}} = \frac{1}{\sqrt{d/2}}\sum_{j=0}^{\frac{d}{2} -1}\hat{a}_{j,0}^{\dagger}\hat{a}_{j,1}^{\dagger}\dots\hat{a}_{j,d-3}^{\dagger}\ket{\text{vac}}~,
\end{equation}
where $\hat{a}_{j,k}^{\dagger}$ creates a photon in the mode $j$ for the $k$-th auxiliary state. Since the state in Eq.(\ref{eq:ancilla_bharos}) has Schmidt rank $2/d^2$, a BSM of all Bell states becomes now achievable with a success probability of $P_{s}=2/d^2$, thus providing an enhancement over previous methods relying on separable auxiliary photons. This work was further extended to odd-prime dimensional systems in \cite{ustun2025fusion}, obtaining a success probability of $P_s = 2/d(d+1)$. In these HD extensions of auxiliary-based protocols, the main concern remains scalability in the generation of entangled auxiliary states, since their complex structure grows with the size of the qudit dimension, making them difficult to engineer. In light of this, in the following we examine nonlinear interactions, highlighting their potential as an alternative route to the implementation of HD BSMs.

\subsubsection{Nonlinear optics}
The theoretical proposal for a BSM for qubits based on a pre-detection squeezing technique, originally introduced in \cite{zaidi2013beating}, has been recently generalized to HD systems in \cite{bianchi2025predetection}. In this work, the success probability was numerically evaluated for dimensions $d = 3, 4, 5$, showing an improvement over previous scaling laws that relied on auxiliary photons \,red{with an amount of squeezing easily achievable within current technology}. Notably, this protocol does not require auxiliary states, as it is provided evidence that their inclusion does not enhance the success probability. The primary limitation of this approach lies in the need for photon-number-resolving detectors.

In \cite{sephton2023quantum}, a nonlinear protocol for teleporting HD spatial information using SFG as a detection mechanism was proposed an experimentally validated. Note that this technique that was already exploited for qubits in \cite{kim2001quantum}. In this scheme, a photon from an entangled pair generated via SPDC, is combined with a spatially encoded coherent state in a second nonlinear crystal. SFG occurs only when their spatial modes are matched, effectively acting as a filter and enabling heralded teleportation. The nonlinear detector functions as a HD, basis-independent spatial-mode filter and supports various encoding schemes, including OAM, Hermite-Gaussian, and arbitrary spatial modes. In addition to providing an experimental proof-of-principle of the above technique, the authors  introduce a theoretical model describing the process as a quantum channel, relating its capacity, that is, the number of modes that can be faithfully transferred, to experimentally accessible parameters like pump beam width. They demonstrate efficient operation up to dimension $d = 14$, laying the groundwork for further research on nonlinear optical BSMs.

Along similar lines, the SFG-based BSM method in \cite{kim2001quantum} has been extended to arbitrary dimensions in \cite{bianchi2025nonlinear}. A key advantage of this generalization is its independence from the choice of encodings. Even if an experimental demonstration of this protocol is still lacking, its performance was evaluated by simulating the fidelity under the presence of crosstalk noise, showing that the realization of a complete BSM is prevented only by the efficiency of the chosen quantum nonlinear process.
Finally, quantum light-matter interactions are also gaining attention in the context of HD states, and seem to offer really promising solutions for the implementation of controlled gates that could overcome the probabilistic limitations of photon-photon interactions \cite{chen2024high, liu2024heralded,tang2025deterministic}.

\subsubsection{Hyper-entanglement}
In \cite{zhang2019arbitrary}, the authors propose a qudit-based scheme for performing a complete BSM. Their method generalizes the concept of hyper-entanglement introduced in \cite{walborn2003hyperentanglement} by utilizing an auxiliary maximally entangled state in a separate Hilbert space—such as path, orbital angular momentum (OAM), or time-bin—to fully resolve the Bell basis in a primary HD space. The authors provide an explicit example of constructing a BSM for angular momentum–path $d=3,4$ systems and outline a methodology to generalize their scheme to arbitrary dimensions.

\section{Continuous-variable Bell state measurements}
\label{sec:CV}
An alternative paradigm for optical quantum information processing is provided by the \textit{continuous-variable} regime~\cite{lloyd1999quantum,braunstein2005quantum}, where information is encoded in the continuous degrees of freedom of the electromagnetic field, known as \emph{quadratures}. For an optical mode described by the bosonic operators $\hat{a}$ and $\hat{a}^\dagger$, they are defined as 
\begin{equation}
    \hat{x} = \frac{1}{2}(\hat{a} + \hat{a}^\dagger), \qquad 
    \hat{p} = \frac{1}{2i}(\hat{a} - \hat{a}^\dagger),
\end{equation}
which satisfy canonical commutation relations and have continuous spectrum. This formalism provides a natural connection between many classes of optical states, such as coherent and
squeezed states, and the Heisenberg uncertainty relations.

Within this framework, \textit{Gaussian states}~\cite{adesso2014continuous} represent the main resource for CV quantum information processing~\cite{lami2018gaussian}, as they can be efficiently generated using linear optics and squeezing operations, and are conveniently described by mean vectors and covariance matrices. However, it has been shown that a genuine quantum advantage often requires the use of \textit{non-Gaussian} resources~\cite{bartlett2002efficient}, sparking growing interest in non-Gaussian state engineering.

A fundamental difference between DV and CV protocols lies in the measurement process. DV measurements typically rely on photon counting, whereas CV protocols use quadrature measurements. The most common example is represented by the so-called \emph{homodyne detection}~\cite{vogel2006quantum}. In this scheme, the signal interferes with a strong coherent local oscillator at a balanced beam splitter, and the difference of the output photocurrents yields the expectation value of the quadrature 
\begin{equation}
    \hat{x}_\phi = \frac{1}{\sqrt{2}}\left(\hat{a}e^{-i\phi} + \hat{a}^\dagger e^{i\phi}\right),
\end{equation}
where $\phi$ is the phase of the local oscillator. By simultaneously measuring two conjugate quadratures via an additional vacuum mode, one obtains \emph{heterodyne detection}, the key ingredient of continuous-variable BSMs.

In the ideal two-mode scenario, a CV BSM corresponds to a projection onto an infinitely squeezed EPR state. Formally, the Bell basis is given by $\ket{\Psi_r} = \hat{D}_{-\hat{r}}\ket{\xi}$, where $\hat{r}=(\hat{x}_1,\hat{p}_1,\hat{x}_2,\hat{p}_2)$ collects the quadratures, $\hat{D}_{\hat{r}}=\exp(i\hat{r}^\mathrm{T}\Omega\hat{r})$ is a displacement operator, and $\Omega=\bigoplus_{i=1}^{2}\Omega_{1}$ is the symplectic form, with $\Omega_{1}=\begin{pmatrix}0&1\\[-2pt]-1&0\end{pmatrix}$. Here, the unnormalized state $\ket{\xi}=\sum_{j=0}^{\infty}\ket{j,j}$ represents a maximally entangled state in the limit of infinite squeezing~\cite{serafini2023quantum}. Experimentally, the Bell projection is implemented by mixing the two modes on a 50:50 beam splitter and measuring the commuting quadrature pair
\begin{equation}
    \hat{x}_- = \hat{x}_1 - \hat{x}_2, \qquad 
    \hat{p}_+ = \hat{p}_1 + \hat{p}_2,
\end{equation}
which deterministically projects onto a continuous entangled basis. Thus, unlike DV Bell measurements, which are fundamentally probabilistic under linear optics, CV BSMs are deterministic.
However, such deterministic behaviour holds only in the ideal case and indeed, when considering finite values of the squeezing parameter, the measured basis deviates from the maximally entangled EPR state, resulting in reduced fidelity for the teleported state and lower-quality entanglement in swapping protocols \cite{hofmann2000fidelity,pirandola2006quantum}. 

After the seminal work \cite{braunstein1998teleportation}, where a CV teleportation protocol based on squeezed states was proposed, the first experiment reporting the unconditional teleportation of a coherent state dates back to \cite{furusawa1998unconditional}, followed by other experimental confirmations few years later \cite{zhang2002quantum,bowen2003experimental}. 
On a similar note, the first theoretical proposal for a CV-based entanglement swapping protocol was provided in \cite{van1999unconditional}, whose experimental proof was reported in \cite{takei2005high}. Remarkably, in this work the authors also report the teleportation of a coherent state with fidelity $F>2/3$, that is beyond the so-called no-cloning limit \cite{grosshans2001quantum}.
More recently, there is a growing interest in hybrid platforms \cite{andersen2015hybrid}, combining discrete- and continuous-variables encodings, both for teleportation \cite{takeda2015entanglement,ulanov2017quantum} and entanglement swapping \cite{lim2016loss,guccione2020connecting} (see also \cite{he2022teleportation} and references therein).

Beyond Gaussian resources, non-Gaussian operations such as photon catalysis, photon addition \cite{duc2024enhancement} and photon subtraction have also attracted considerable attention, providing numerous evidences of enhanced performances in quantum teleportation (see \cite{arora2025continuous}). In light of the recent interest in non-Gaussian resources, exploring their integration with CV BSMs appears to be a promising avenue for future research.

\section{Applications}
\label{sec:app}
\begin{figure}
    \centering
    \includegraphics[width=0.5\linewidth]{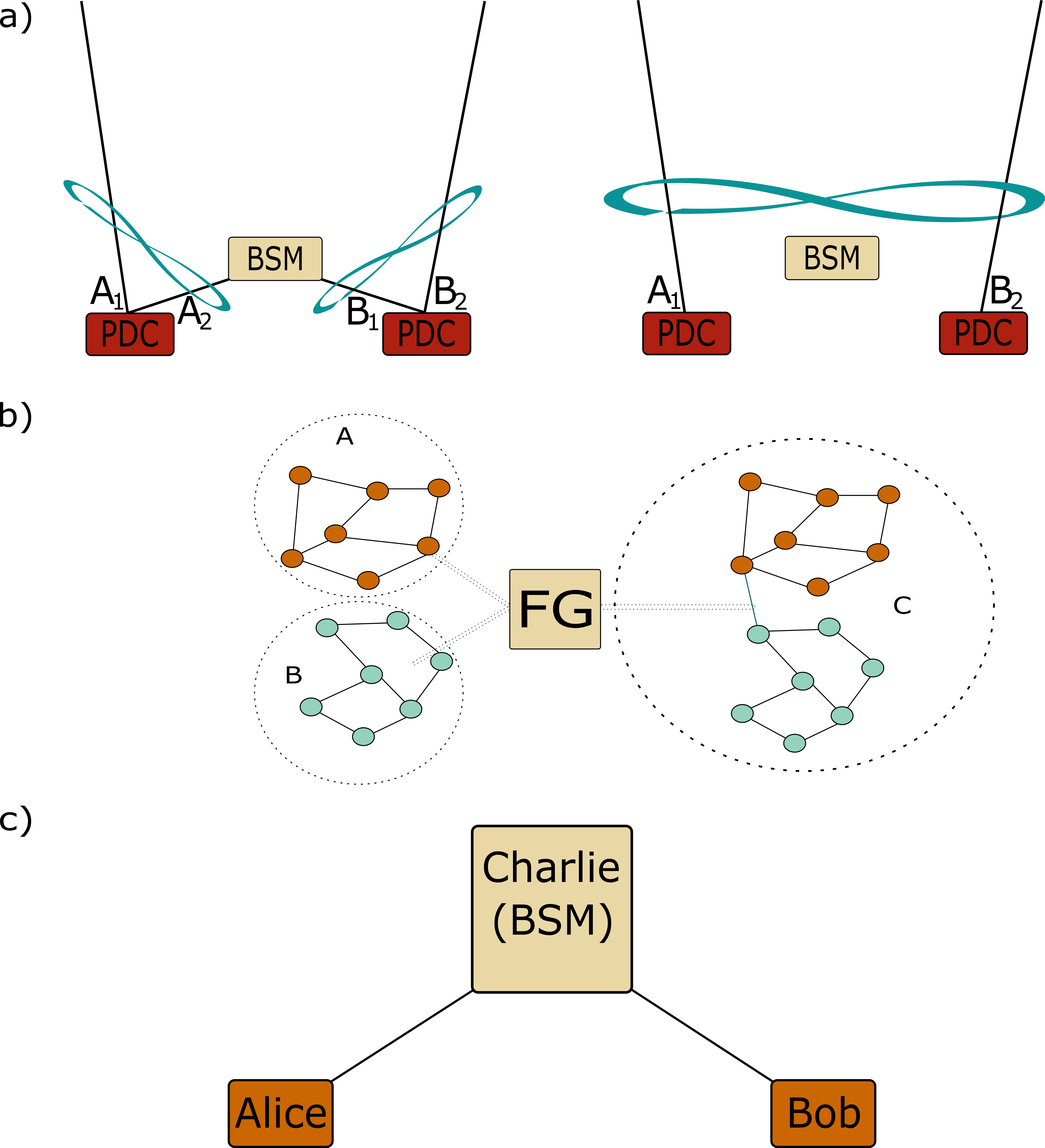}
    \caption{The most common applications of BSMs in quantum information: a) Entanglement swapping protocol, which allows to share an entangled photon pair between distant parties that never interacted before; b) Fusion gates for measurement-based quantum computation, where large cluster states — used as a resource for quantum operations via measurement protocols — are built by entangling initially uncorrelated photons; c) Measurement-device-independent quantum-key-distribution, in which an intermediate BSM removes the need for trusted nodes, enabling end-to-end security even when the measurement stations are untrusted.}
    \label{fig:app}
\end{figure}

As previously mentioned, BSMs are a fundamental component of many quantum information protocols. In this section, we focus on three key applications: quantum repeaters, fusion-based quantum computation and quantum key distribution, which are schematically displayed in Fig.~\ref{fig:app}. For each application, we review the current state of the art and discuss the main challenges in their implementations. 

\subsection{Quantum repeaters}
Quantum repeaters are essential for enabling long-distance quantum communication by overcoming the limitations imposed by photon loss and noise in optical channels \cite{pirandola2017fundamental}. Unlike classical signal amplifiers, quantum repeaters cannot clone quantum states due to the no-cloning theorem \cite{wootters1982single}. Instead, they utilize techniques such as entanglement distribution \cite{Distribution}, purification \cite{purification}, and swapping \cite{Pan1998} to extend the range of quantum networks while preserving quantum coherence. The basic principle involves dividing the communication channel into shorter segments, generating entanglement between neighboring nodes, and then performing entanglement swapping to establish end-to-end entanglement \cite{briegel1998quantum}.

The development of quantum repeaters has progressed through several generations, each improving performance and scalability. For an extensive review on quantum repeaters, see \cite{azuma2023quantum}. Here, we focus solely on the key points where BSMs play a role, without addressing other important aspects of quantum repeater research, such as quantum error correction and entanglement distillation protocols. First-generation repeaters rely on entanglement swapping combined with probabilistic entanglement purification \cite{dur1999quantum}. While these systems are close to experimental feasibility using quantum memories leveraging light-matter interactions \cite{van2006hybrid, ladd2006hybrid, bergmann2019hybrid},  based on trapped ions (see e.g. \cite{sangouard2009quantum, krutyanskiy2023telecom}) or atomic ensembles \cite{chen2007fault, sangouard2008robust, han2010quantum, sangouard2011quantum}, they are limited by finite memory coherence times and the resource demands of purification. Second-generation repeaters \cite{jiang2009quantum} incorporate quantum error correction to protect against operational errors, enhancing noise robustness \cite{muralidharan2014ultrafast}. These require high-fidelity gates and efficient error-correcting codes, which present significant experimental challenges. Third-generation repeaters (see, e.g. \cite{fowler2010surface}), aim for full fault tolerance through topological error correction and deterministic entanglement swapping \cite{muralidharan2016optimal}, demanding breakthroughs in hardware and control.  

Several major challenges remain in realizing practical quantum repeaters. Quantum memories must preserve coherence long enough to complete entanglement distribution across network segments, often requiring millisecond-to-second storage times for long distances \cite{simon2010quantum}. Photon loss in optical fibers and detector noise degrade performance, necessitating advanced error mitigation. On the CV side, many proposals were designed for quantum repeaters which rely on the deterministic nature of BSMs \cite{polkinghorne1999continuous, ide2001continuous, hoelscher2011optimal, hofer2013time, dias2020quantum, chen2020parametric, liu2022all}. Recent advances include hybrid approaches combining DV and CV systems to enhance efficiency \cite{takeda2015entanglement,guccione2020connecting}. All-photonic repeaters \cite{azuma2015all} and their variants \cite{pant2017rate,ewert2016ultrafast,lee2019fundamental}, instead, eliminate the need for quantum memories by using graph states \cite{li2019experimental, shchukin2021waiting} (see Sec. \ref{sec: fusion} for additional references on measurement-based quantum computing). The key advantage consists of a workaround of the online, probabilistic, single-shot BSM. Furthermore, universality is not required and the resource state may be prepared offline, reducing waiting times that may increase photon losses. The structure of the graph state and the optimal choice of error-correcting code depend on the specific repeater model, and the most suitable repeater architecture to adopt remains an open question (see, e.g., \cite{zwerger2016measurement} for an alternative scheme of a photonic quantum repeater). Other proposals include CV \cite{fukui2021all} and hybrid DV-CV \cite{rozpedek2021quantum}. In general, a cluster state is shared between two repeater nodes, with each station receiving halves of two different cluster states. A fixed number of BSMs are performed, and based on their outcomes, measurements on the new cluster are used to distill Bell pairs. Performing more BSMs increases the probability of forming larger clusters that entangle distant nodes.

\subsection{Fusion-based quantum computation}
\label{sec: fusion}
BSMs also play a central role in the photonic implementation of measurement-based quantum computation \cite{raussendorf2001one}, known as fusion-based quantum computation \cite{fusion1,fusion2,fusion3}, which stands out as a promising paradigm for scalable photonic quantum computing. This approach leverages entangled resource states and adaptive measurements to perform universal quantum operations \cite{stefano1}. Conceptually, gate-based and fusion-based models reflect the duality between the Schr\"{o}dinger and the Heisenberg pictures, with the latter differing in the preparation of resources, which is carried out at the very beginning of the computation. This typically involves constructing large entangled states by fusing smaller states, such as Bell pairs or graph states. The fusion gate itself is essentially a BSM, which can be performed either probabilistically or deterministically, depending on whether the information is encoded in DV or CV.
Specifically, in the discrete case, the HOM effect at the output ports of a beam splitter provides a key mechanism to generate photonic graph states, acting as an entangling gate that fuses particles together. Hence, fusion gates are probabilistic, but they can be implemented offline and their efficiency can still be enhanced using the techniques described in Sections \ref{sec:BSM-qubit} and \ref{sec:BSM-HD}. This approach is particularly well suited to photonic systems \cite{stefano2, stefano5}, where the implementation of a deterministic entangling gate can be performed before the computation. Furthermore, a major advantage of fusion-based quantum computing is its intrinsic tolerance to photon loss and operational errors \cite{stefano4}. Nonetheless, high-quality single-photon sources and detectors still remain an essential requirement. Recent experiments have demonstrated small-scale fusion gates using linear optics, marking important steps toward fault-tolerant, scalable architectures \cite{song2024encoded,psiquantum2025manufacturable}.

Cluster states are employed as a resource also in CV quantum information processing (see, e.g. \cite{menicucci2006universal, gu2009quantum, menicucci2014fault}). Many techniques and protocols take advantage of the Gaussian and deterministic nature of homodyne detection \cite{booth2023flow, du2025complete}.
However, CV fusion-based computation is limited by imperfections arising from finite squeezing, which introduce computational errors when measuring CV cluster states. For a concise overview of CV fusion-based schemes and an analysis on squeezing imperfections, we refer the interested reader to \cite{xin2023qumode}.

\subsection{Quantum key distribution}
QKD \cite{pirandola2020advances,xu2020secure} is a technique that enables the secure exchange of symmetric cryptographic keys between distant parties by exploiting the fundamental principles of quantum mechanics. Thanks to rapid technological progress, QKD has become one of the most mature and promising applications of quantum information science, with successful implementations achieving increasingly long transmission distances through both free-space \cite{aspelmeyer2003long,vallone2015experimental,liao2017satellite,cocchi2025time} and optical fiber channels \cite{ding2017high,chen2022quantum,mueller2025performance,ribezzo2023deploying}

However, a major challenge lies in the gap between the theoretical models of QKD protocols, often based on idealized assumptions, and their practical realizations. Discrepancies between theory and implementation can give rise to security loopholes, highlighting the need for frameworks that are robust against device imperfections. A significant advancement in this direction is provided by measurement-device-independent (MDI) QKD \cite{lo2012measurement, xu2015discrete}, together with its recent version, twin-field (TF) QKD \cite{lucamarini2018overcoming}, both of which eliminate the need to trust measurement devices and thus mitigate vulnerabilities to so-called side-channel attacks \cite{lydersen2010hacking,xu2013practical,jo2019enhanced}.

In both MDI and TF QKD, the two legitimate parties (commonly referred to as Alice and Bob) send their encoded quantum states to an untrusted third party who performs a BSM. The outcome of this measurement enables the two parties to establish a shared key without requiring trust in the measurement process itself. Experimental demonstrations of these protocols have been successfully carried out in various platforms, both for MDI QKD \cite{tang2014measurement,tang2016measurement,yin2016measurement,wang2021measurement} and for TF QKD (see \cite{wang2022twin} and references therein).

While MDI and TF QKD are compatible with current technological capabilities, they require a well-characterized and trusted quantum transmitter. A more comprehensive approach that also removes assumptions on the source is offered by device-independent (DI) QKD \cite{mayers1998quantum, zapatero2023advances}, in which entangled photon pairs from an untrusted source are distributed to the users. In this setting, the security of the protocol is certified by the violation of a Bell inequality \cite{ekert1991quantum}. However, a key assumption in DI QKD is the ability to close all relevant loopholes in a Bell test—an experimentally demanding task that has only recently been demonstrated \cite{nadlinger2022experimental,zhang2022device,liu2022toward}. In this context, BSMs play a crucial role by enabling heralding mechanisms that identify successful entanglement distribution events without introducing detection loopholes (see, e.g., \cite{zapatero2023advances}).

\section{Conclusions}
\label{sec:conclusions}
In this review, we provided a comprehensive overview of Bell state measurements, examining both their foundations as well as their most recent theoretical proposals and experimental implementations.
Our discussion covered Bell state measurements in qubit systems, where they are best understood and most widely realized, as well as in high-dimensional systems, which offer enhanced capabilities but also introduce additional experimental challenges. We analyzed the fundamental limitations that hinder the realization of highly efficient Bell state measurements in quantum optical platforms.
In particular, for qubit systems, the restriction to linear optics results in intrinsically probabilistic Bell measurements, while for qudit systems encoded in discrete variables, this limitation is even more severe, as unambiguous Bell state discrimination is fundamentally impossible.
We therefore surveyed the main strategies developed to overcome these constraints, including the use of auxiliary photons, nonlinear optical interactions and hyper-entanglement.
Since continuous-variable encoding can, in principle, enable deterministic Bell measurements and is therefore not subject to the aforementioned limitations, we primarily focused on discrete-variable systems.
Nevertheless, we also considered continuous-variable systems, in which the achievable fidelity is constrained by the finite level of squeezing attainable in realistic experimental settings. As a result, the choice between discrete- and continuous-variable encodings depends strongly on the specific protocol and operational requirements.
With this perspective, we reviewed the key applications of Bell state measurements in several quantum information protocols, including quantum repeaters, quantum key distribution and fusion-based quantum computation. Notably, all these topics are of fundamental importance for the realization of a quantum internet, which represents a major goal of the quantum information community.

We conclude this review by identifying several directions for future research.
As we have discussed, the two main research lines in Bell measurements are auxiliary photons and nonlinear optics. For qubit systems, the scaling of the success probability in the former schemes depends on the number of auxiliary states, which are usually highly entangled, and modes employed. Conversely, the latter approaches typically make the protocol deterministic, with the trade-off lying in the efficiency of the experimental implementation.

For high-dimensional systems, however, the success probability of auxiliary methods is further hindered by the dimension of the Hilbert space, a limitation that is, in principle, not shared by nonlinear techniques. Based on this observation, from a purely theoretical perspective, we expect nonlinear interactions to offer the most promising alternative for the implementation of efficient high-dimensional Bell-measurement protocols. Nevertheless, we stress that even on the theoretical side, a comprehensive description of higher-order nonlinear effects and their application to Bell measurements is still lacking. 
In this direction, we highlight several aspects that may be worth investigating:
\begin{itemize}
    \item Recent experiments \cite{guccione2020connecting} have demonstrated the potential of hybrid approaches that combine deterministic continuous-variables operations with the robustness of discrete-variables photonic. Moreover, the rapid progress in light-matter interfaces --including atomic ensembles, solid-state emitters, and integrated photonic–quantum-memory platforms-- opens the door to deterministic entangling operations and enhanced Bell measurements. Realizing scalable and low-loss hybrid modules represents an important open challenge for the implementation of reliable quantum networks.
    \item Extending the results of \cite{calsamiglia2002generalized} to scenarios involving higher-order nonlinearities could yield new insights into the discrimination of Bell states using physical resources beyond the linear regime. Developing generalized no-go theorems or identifying minimal nonlinear interactions required to surpass linear-optical bounds represents an important theoretical challenge. However, one should keep in mind that at given scales, nonlinear optics becomes inherently multimode \cite{shapiro2006single}. Although this fact might appear as a limitation for the single-mode description of the underlying physics, it may nevertheless open the opportunity to develop new protocols in alternative quantum optical regimes \cite{yanagimoto2024mesoscopic}.
    \item To the best of our knowledge, protocols where both auxiliary states and weak nonlinearities are exploited have not been explored yet. Such an approach could unlock enhanced performance while avoiding the substantial experimental overhead associated with strong nonlinear interactions. Early numerical results, such as those reported in \cite{bianchi2025predetection}, showed how combining squeezing together with single-photon auxiliary states does not allow for a further boost in the success probability. Nevertheless, a more comprehensive analysis is needed to provide a better understanding of the interplay between these two approaches.
    \item The nonlinear effects discussed throughout this review naturally become more pronounced for input states with higher energy or photon number (see, e.g., \cite{sephton2023quantum}). This observation points towards the development of new strategies, specifically tailored to benefit from nonlinear interactions. The design of such protocols, including realistic noise modeling and experimental feasibility studies, remains an open opportunity.
    \item Non-Gaussian operations, such as photon addition and photon subtraction, as well as sum-frequency generation, four-wave mixing, and other higher-order nonlinear processes, appear to be a promising route for future research. In light of the recent interest in non-Gaussian resources, we believe that a systematic analysis of Bell state measurement protocols based on such resources could lead to substantially improved performances.
    \end{itemize}

\section{Data availability statement}
All data that support the findings of this study are included within the article.

\section{Conflict of interest statement}
The authors declare that the research was conducted in the absence of any commercial or financial relationships that could be construed as a potential conflict of interest.

\section{Acknowledgments}

LB thanks Francesco Salusti and Giorgio De Pascalis for fruitful discussions.
LB and DB acknowledge support from the European Union ERC StG, QOMUNE, 101077917. CM acknowledges support from the European Union - NextGeneration EU, "Integrated infrastructure initiative in Photonic and Quantum Sciences" - I-PHOQS [IR0000016, ID D2B8D520, CUP B53C22001750006]. 

We acknowledge that despite our best efforts to provide a comprehensive list of references, some relevant works may have been inadvertently omitted. We sincerely apologize for any such oversights and extend our appreciation to all contributors to this active and rapidly evolving field.

\bibliography{biblio}

@article{erhard2020advances,
  title={Advances in high-dimensional quantum entanglement},
  author={Erhard, Manuel and Krenn, Mario and Zeilinger, Anton},
  journal={Nature Reviews Physics},
  volume={2},
  number={7},
  pages={365--381},
  year={2020},
  publisher={Nature Publishing Group UK London},
doi={ https://doi.org/10.1038/s42254-020-0193-5}
}

@article{browne2005resource,
  title={Resource-efficient linear optical quantum computation},
  author={Browne, Daniel E and Rudolph, Terry},
  journal={Physical Review Letters},
  volume={95},
  number={1},
  pages={010501},
  year={2005},
  publisher={APS},
  doi={10.1103/PhysRevLett.95.010501}
}

@article{duvsek2001discrimination,
  title={Discrimination of the Bell states of qudits by means of linear optics},
  author={Du{\v{s}}ek, Miloslav},
  journal={Optics communications},
  volume={199},
  number={1-4},
  pages={161--166},
  year={2001},
  publisher={Elsevier},
doi={
https://doi.org/10.1016/S0030-4018%2801%2901565-6}
}

@article{mair2001entanglement,
  title={Entanglement of the orbital angular momentum states of photons},
  author={Mair, Alois and Vaziri, Alipasha and Weihs, Gregor and Zeilinger, Anton},
  journal={Nature},
  volume={412},
  number={6844},
  pages={313--316},
  year={2001},
  publisher={Nature Publishing Group UK London},
 doi ={https://doi.org/10.1038/35085529}
}

@article{mclaren2012entangled,
  title={Entangled Bessel-Gaussian beams},
  author={McLaren, Melanie and Agnew, Megan and Leach, Jonathan and Roux, Filippus S and Padgett, Miles J and Boyd, Robert W and Forbes, Andrew},
  journal={Optics express},
  volume={20},
  number={21},
  pages={23589--23597},
  year={2012},
  publisher={Optical Society of America},
 doi = {https://doi.org/10.1364/OE.20.023589}
}

@article{spagnolo2014experimental,
  title={Experimental Boson Sampling with integrated photonics},
  author={Spagnolo, N and Vitelli, C and Bentivegna, M and Flamini, F and Mataloni, P and Sciarrino, F and Brod, Daniel J and Galv{\~a}o, Ernesto F and Crespi, Andrea and Ramponi, Roberta and others},
  journal={Quantum Information and Measurement},
  year={2014},
  publisher={Optica Publishing Group},
  doi ={https://doi.org/10.1364/QIM.2014.QTh1A.3}
}

@article{bayerbach2023bell,
  title={Bell-state measurement exceeding 50\% success probability with linear optics},
  author={Bayerbach, Matthias J and D’Aurelio, Simone E and Van Loock, Peter and Barz, Stefanie},
  journal={Science Advances},
  volume={9},
  number={32},
  pages={eadf4080},
  year={2023},
  publisher={American Association for the Advancement of Science},
doi={10.1126/sciadv.adf4080}
}

@article{sych2009complete,
  title={A complete basis of generalized Bell states},
  author={Sych, Denis and Leuchs, Gerd},
  journal={New Journal of Physics},
  volume={11},
  number={1},
  pages={013006},
  year={2009},
  publisher={IOP Publishing},
doi={10.1088/1367-2630/11/1/013006}
}

@article{d2025boosted,
  title={Boosted quantum teleportation},
  author={D’Aurelio, Simone E and Bayerbach, Matthias J and Barz, Stefanie},
  journal={npj Quantum Information},
  volume={11},
  number={1},
  pages={37},
  year={2025},
  publisher={Nature Publishing Group UK London},
doi={https://doi.org/10.1038/s41534-025-00992-4}
}

@article{da2021path,
  title={Path-encoded high-dimensional quantum communication over a 2-km multicore fiber},
  author={Da Lio, Beatrice and Cozzolino, Daniele and Biagi, Nicola and Ding, Yunhong and Rottwitt, Karsten and Zavatta, Alessandro and Bacco, Davide and Oxenl{\o}we, Leif K},
  journal={npj Quantum Information},
  volume={7},
  number={1},
  pages={63},
  year={2021},
  publisher={Nature Publishing Group UK London},
 doi={https://doi.org/10.1038/s41534-021-00398-y}
}

@article{ma2025preparation,
  title={Preparation of heralded high-fidelity high-dimensional entangled states implemented with qudit-encoded photon systems},
  author={Ma, Ming and Tan, Qiu-Lin and Du, Fang-Fang},
  journal={Optics Express},
  volume={33},
  number={11},
  pages={23678--23691},
  year={2025},
  publisher={Optica Publishing Group},
  doi={https://doi.org/10.1364/OE.561002}
}

@article{brecht2015photon,
  title={Photon temporal modes: a complete framework for quantum information science},
  author={Brecht, Benjamin and Reddy, Dileep V and Silberhorn, Christine and Raymer, Michael G},
  journal={Physical Review X},
  volume={5},
  number={4},
  pages={041017},
  year={2015},
  publisher={APS},
  doi= {https://doi.org/10.1103/PhysRevX.5.041017}
}

@article{yu2025quantum,
  title={Quantum key distribution implemented with d-level time-bin entangled photons},
  author={Yu, Hao and Sciara, Stefania and Chemnitz, Mario and Montaut, Nicola and Crockett, Benjamin and Fischer, Bennet and Helsten, Robin and Wetzel, Benjamin and Goebel, Thorsten A and Kr{\"a}mer, Ria G and others},
  journal={Nature Communications},
  volume={16},
  number={1},
  pages={171},
  year={2025},
  publisher={Nature Publishing Group UK London},
 doi ={https://doi.org/10.1038/s41467-024-55345-0}
}

@article{ralph2002linear,
  title={Linear optical controlled-NOT gate in the coincidence basis},
  author={Ralph, Timothy C and Langford, Nathan K and Bell, TB and White, AG},
  journal={Physical Review A},
  volume={65},
  number={6},
  pages={062324},
  year={2002},
  publisher={APS},
doi={https://doi.org/10.1103/PhysRevA.65.062324}
}

@article{lucamarini2018overcoming,
  title={Overcoming the rate--distance limit of quantum key distribution without quantum repeaters},
  author={Lucamarini, Marco and Yuan, Zhiliang L and Dynes, James F and Shields, Andrew J},
  journal={Nature},
  volume={557},
  number={7705},
  pages={400--403},
  year={2018},
  publisher={Nature Publishing Group UK London},
doi={https://doi.org/10.1038/s41586-018-0066-6}
}

@book{bertlmann2002bell,
  title={Bell’s theorem, information and quantum physics},
  author={Bertlmann, Reinhold A and Zeilinger, Anton},
  year={2002},
  publisher={Springer},
 doi={https://doi.org/10.1007/978-3-662-05032-3_17}
}

@article{weinfurter1994experimental,
  title={Experimental Bell-state analysis},
  author={Weinfurter, Harald},
  journal={Europhysics Letters},
  volume={25},
  number={8},
  pages={559},
  year={1994},
  publisher={IOP Publishing},
  doi = {10.1209/0295-5075/25/8/001}
}

@article{guccione2020connecting,
  title={Connecting heterogeneous quantum networks by hybrid entanglement swapping},
  author={Guccione, Giovanni and Darras, Tom and Le Jeannic, Hanna and Verma, Varun B and Nam, Sae Woo and Cavaill{\`e}s, Adrien and Laurat, Julien},
  journal={Science advances},
  volume={6},
  number={22},
  pages={eaba4508},
  year={2020},
  publisher={American Association for the Advancement of Science},
  doi={10.1126/sciadv.aba4508}
}

@article{shapiro2006single,
  title={Single-photon Kerr nonlinearities do not help quantum computation},
  author={Shapiro, Jeffrey H},
  journal={Physical Review A—Atomic, Molecular, and Optical Physics},
  volume={73},
  number={6},
  pages={062305},
  year={2006},
  publisher={APS},
  doi={10.1103/PhysRevA.73.062305}
}

@article{arora2025continuous,
  title={Continuous-variable quantum teleportation using a photon-subtracted and photon-added two-mode squeezed coherent state},
  author={Arora, Shikhar and Kumar, Chandan and Arvind},
  journal={Physical Review A},
  volume={111},
  number={2},
  pages={022402},
  year={2025},
  publisher={APS},
  doi={10.1103/PhysRevA.111.022402}
}

@article{ulanov2017quantum,
  title={Quantum teleportation between discrete and continuous encodings of an optical qubit},
  author={Ulanov, Alexander E and Sychev, Demid and Pushkina, Anastasia A and Fedorov, Ilya A and Lvovsky, AI},
  journal={Physical review letters},
  volume={118},
  number={16},
  pages={160501},
  year={2017},
  publisher={APS},
  doi={10.1103/PhysRevLett.118.160501}
}

@article{lim2016loss,
  title={Loss-resilient photonic entanglement swapping using optical hybrid states},
  author={Lim, Youngrong and Joo, Jaewoo and Spiller, Timothy P and Jeong, Hyunseok},
  journal={Physical Review A},
  volume={94},
  number={6},
  pages={062337},
  year={2016},
  publisher={APS},
  doi={10.1103/PhysRevA.94.062337}
}

@article{he2022teleportation,
  title={Teleportation of hybrid entangled states with continuous-variable entanglement},
  author={He, Mingjian and Malaney, Robert},
  journal={Scientific Reports},
  volume={12},
  number={1},
  pages={17169},
  year={2022},
  publisher={Nature Publishing Group UK London},
  doi={10.1038/s41598-022-21283-4}
}

@article{pirandola2015advances,
  title={Advances in quantum teleportation},
  author={Pirandola, Stefano and Eisert, Jens and Weedbrook, Christian and Furusawa, Akira and Braunstein, Samuel L},
  journal={Nature photonics},
  volume={9},
  number={10},
  pages={641--652},
  year={2015},
  publisher={Nature Publishing Group UK London},
  doi={10.1038/nphoton.2015.154}
}

@article{pan1998experimental,
  title={Experimental entanglement swapping: entangling photons that never interacted},
  author={Pan, Jian-Wei and Bouwmeester, Dik and Weinfurter, Harald and Zeilinger, Anton},
  journal={Physical review letters},
  volume={80},
  number={18},
  pages={3891},
  year={1998},
  publisher={APS},
  doi={10.1103/PhysRevLett.80.3891}
}

@article{boschi1998experimental,
  title={Experimental realization of teleporting an unknown pure quantum state via dual classical and Einstein-Podolsky-Rosen channels},
  author={Boschi, Danilo and Branca, Salvatore and De Martini, Francesco and Hardy, Lucien and Popescu, Sandu},
  journal={Physical Review Letters},
  volume={80},
  number={6},
  pages={1121},
  year={1998},
  publisher={APS},
  doi={10.1103/PhysRevLett.80.1121}
}

@article{wein2016efficiency,
  title={Efficiency of an enhanced linear optical Bell-state measurement scheme with realistic imperfections},
  author={Wein, Stephen and Heshami, Khabat and Fuchs, Christopher A and Krovi, Hari and Dutton, Zachary and Tittel, Wolfgang and Simon, Christoph},
  journal={Physical Review A},
  volume={94},
  number={3},
  pages={032332},
  year={2016},
  publisher={APS},
  doi={10.1103/PhysRevA.94.032332}
}

@article{bouwmeester1997experimental,
  title={Experimental quantum teleportation},
  author={Bouwmeester, Dik and Pan, Jian-Wei and Mattle, Klaus and Eibl, Manfred and Weinfurter, Harald and Zeilinger, Anton},
  journal={Nature},
  volume={390},
  number={6660},
  pages={575--579},
  year={1997},
  publisher={Nature Publishing Group UK London},
  doi ={https://doi.org/10.1038/37539}
}

@article{braunstein1995measurement,
  title={Measurement of the Bell operator and quantum teleportation},
  author={Braunstein, Samuel L and Mann, Ady},
  journal={Physical Review A},
  volume={51},
  number={3},
  pages={R1727},
  year={1995},
  publisher={APS},
  doi={https://doi.org/10.1103/PhysRevA.51.R1727}
}

@article{carollo2001role,
  title={The role of auxiliary states in state discrimination with linear optical devices},
  author={Carollo, Angelo and Palma, G Massimo},
  journal={arXiv preprint quant-ph/0106041},
  year={2001},
doi={ https://doi.org/10.1080/09500340110098370}
}

@article{ewert20143,
  title={3/4-efficient bell measurement with passive linear optics and unentangled ancillae},
  author={Ewert, Fabian and van Loock, Peter},
  journal={Physical review letters},
  volume={113},
  number={14},
  pages={140403},
  year={2014},
  publisher={APS},
doi={https://doi.org/10.1103/PhysRevLett.113.140403}
}

@article{grice2011arbitrarily,
  title={Arbitrarily complete Bell-state measurement using only linear optical elements},
  author={Grice, Warren P},
  journal={Physical Review A—Atomic, Molecular, and Optical Physics},
  volume={84},
  number={4},
  pages={042331},
  year={2011},
  publisher={APS},
doi={https://doi.org/10.1103/PhysRevA.84.042331}
}

@article{zhang2019arbitrary,
  title={Arbitrary two-particle high-dimensional Bell-state measurement by auxiliary entanglement},
  author={Zhang, Hao and Zhang, Chao and Hu, Xiao-Min and Liu, Bi-Heng and Huang, Yun-Feng and Li, Chuan-Feng and Guo, Guang-Can},
  journal={Physical Review A},
  volume={99},
  number={5},
  pages={052301},
  year={2019},
  publisher={APS},
doi={https://doi.org/10.1103/PhysRevA.99.052301}
}

@book{nielsen2001quantum,
  title={Quantum computation and quantum information},
  author={Nielsen, Michael A and Chuang, Isaac L},
  volume={2},
  year={2001},
  publisher={Cambridge university press Cambridge},
  doi={https://doi.org/10.1017/CBO9780511976667}
}

@article{walborn2003hyperentanglement,
  title={Hyperentanglement-assisted Bell-state analysis},
  author={Walborn, SP and P{\'a}dua, S and Monken, CH},
  journal={Physical Review A},
  volume={68},
  number={4},
  pages={042313},
  year={2003},
  publisher={APS},
doi={https://doi.org/10.1103/PhysRevA.68.042313}
}

@article{olivo2018investigating,
  title={Investigating the optimality of ancilla-assisted linear optical Bell measurements},
  author={Olivo, Andrea and Grosshans, Fr{\'e}d{\'e}ric},
  journal={arXiv preprint arXiv:1806.01243},
  year={2018},
doi={
https://doi.org/10.1103/PhysRevA.98.042323}
}

@article{grosshans2001quantum,
  title={Quantum cloning and teleportation criteria for continuous quantum variables},
  author={Grosshans, Fr{\'e}d{\'e}ric and Grangier, Philippe},
  journal={Physical Review A},
  volume={64},
  number={1},
  pages={010301},
  year={2001},
  publisher={APS},
  doi={10.1103/PhysRevA.64.010301}
}

@article{yanagimoto2024mesoscopic,
  title={Mesoscopic ultrafast nonlinear optics—the emergence of multimode quantum non-Gaussian physics},
  author={Yanagimoto, Ryotatsu and Ng, Edwin and Jankowski, Marc and Nehra, Rajveer and McKenna, Timothy P and Onodera, Tatsuhiro and Wright, Logan G and Hamerly, Ryan and Marandi, Alireza and Fejer, M\_M and others},
  journal={Optica},
  volume={11},
  number={7},
  pages={896--918},
  year={2024},
  publisher={Optica Publishing Group},
 doi = {https://doi.org/10.1364/OPTICA.514075}
}

@article{lutkenhaus1999bell,
  title={Bell measurements for teleportation},
  author={L{\"u}tkenhaus, Norbert and Calsamiglia, John and Suominen, K-A},
  journal={Physical Review A},
  volume={59},
  number={5},
  pages={3295},
  year={1999},
  publisher={APS},
doi={https://doi.org/10.1103/PhysRevA.59.3295}
}

@article{calsamiglia2002generalized,
  title={Generalized measurements by linear elements},
  author={Calsamiglia, John},
  journal={Physical Review A},
  volume={65},
  number={3},
  pages={030301},
  year={2002},
  publisher={APS},
doi={https://doi.org/10.1103/PhysRevA.65.030301}
}

@article{kilmer2019boosting,
  title={Boosting linear-optical Bell measurement success probability with predetection squeezing and imperfect photon-number-resolving detectors},
  author={Kilmer, Thomas and Guha, Saikat},
  journal={Physical Review A},
  volume={99},
  number={3},
  pages={032302},
  year={2019},
  publisher={APS},
doi={ https://doi.org/10.1103/PhysRevA.99.032302}
}

@article{hauser2025boosted,
  title={Boosted Bell-state measurements for photonic quantum computation},
  author={Hauser, Nico and Bayerbach, Matthias J and D’Aurelio, Simone E and Weber, Raphael and Santandrea, Matteo and Kumar, Shreya P and Dhand, Ish and Barz, Stefanie},
  journal={npj Quantum Information},
  volume={11},
  number={1},
  pages={41},
  year={2025},
  publisher={Nature Publishing Group UK London},
doi = {https://doi.org/10.1038/s41534-025-00986-2}
}

@article{van1999unconditional,
  title={Unconditional teleportation of continuous-variable entanglement},
  author={Van Loock, Peter and Braunstein, Samuel L},
  journal={Physical Review A},
  volume={61},
  number={1},
  pages={010302},
  year={1999},
  publisher={APS},
  doi={10.1103/PhysRevA.61.010302}
}

@article{takeda2015entanglement,
  title={Entanglement swapping between discrete and continuous variables},
  author={Takeda, Shuntaro and Fuwa, Maria and van Loock, Peter and Furusawa, Akira},
  journal={Physical review letters},
  volume={114},
  number={10},
  pages={100501},
  year={2015},
  publisher={APS},
  doi={10.1103/PhysRevLett.114.100501}
}

@article{luo2019quantum,
  title={Quantum teleportation in high dimensions},
  author={Luo, Yi-Han and Zhong, Han-Sen and Erhard, Manuel and Wang, Xi-Lin and Peng, Li-Chao and Krenn, Mario and Jiang, Xiao and Li, Li and Liu, Nai-Le and Lu, Chao-Yang and others},
  journal={Physical review letters},
  volume={123},
  number={7},
  pages={070505},
  year={2019},
  publisher={APS},
doi={https://doi.org/10.1103/PhysRevLett.123.070505}
}

@article{knill2001scheme,
  title={A scheme for efficient quantum computation with linear optics},
  author={Knill, Emanuel and Laflamme, Raymond and Milburn, Gerald J},
  journal={nature},
  volume={409},
  number={6816},
  pages={46--52},
  year={2001},
  publisher={Nature Publishing Group UK London},
  doi = {https://doi.org/10.1038/35051009}
}

@article{bennett1993teleporting,
  title={Teleporting an unknown quantum state via dual classical and Einstein-Podolsky-Rosen channels},
  author={Bennett, Charles H and Brassard, Gilles and Cr{\'e}peau, Claude and Jozsa, Richard and Peres, Asher and Wootters, William K},
  journal={Physical review letters},
  volume={70},
  number={13},
  pages={1895},
  year={1993},
  publisher={APS},
doi={https://doi.org/10.1103/PhysRevLett.70.1895}
}

@article{kim2001quantum,
  title={Quantum teleportation of a polarization state with a complete Bell state measurement},
  author={Kim, Yoon-Ho and Kulik, Sergei P and Shih, Yanhua},
  journal={Physical Review Letters},
  volume={86},
  number={7},
  pages={1370},
  year={2001},
  publisher={APS},
doi={https://doi.org/10.1103/PhysRevLett.86.1370}
}

@article{yamazaki2025linear,
  title={Linear-optical fusion boosted by high-dimensional entanglement},
  author={Yamazaki, Tomohiro and Azuma, Koji},
  journal={Physical Review Letters},
  volume={134},
  number={20},
  pages={200801},
  year={2025},
  publisher={APS},
 doi = {https://doi.org/10.1103/PhysRevLett.134.200801}
}

@article{zukowski1997realizable,
  title={Realizable higher-dimensional two-particle entanglements via multiport beam splitters},
  author={{\.Z}ukowski, Marek and Zeilinger, Anton and Horne, Michael A},
  journal={Physical Review A},
  volume={55},
  number={4},
  pages={2564},
  year={1997},
  publisher={APS},
doi={https://doi.org/10.1103/PhysRevA.55.2564}
}

@article{bharos2024efficient,
  title={Efficient high-dimensional entangled state analyzer with linear optics},
  author={Bharos, Niv and Markovich, Liubov and Borregaard, Johannes},
  journal={Quantum},
  volume={9},
  pages={1711},
  year={2025},
  publisher={Verein zur F{\"o}rderung des Open Access Publizierens in den Quantenwissenschaften},
 doi = {https://doi.org/10.22331/q-2025-04-18-1711}
}

@article{zaidi2013beating,
  title={Beating the one-half limit of ancilla-free linear optics Bell measurements},
  author={Zaidi, Hussain A and van Loock, Peter},
  journal={Physical review letters},
  volume={110},
  number={26},
  pages={260501},
  year={2013},
  publisher={APS},
doi={https://doi.org/10.1103/PhysRevLett.110.260501}
}

@article{bianchi2025predetection,
  title={Predetection squeezing as a resource for high-dimensional Bell-state measurements},
  author={Bianchi, Luca and Marconi, Carlo and Sperling, Jan and Bacco, Davide},
  journal={Physical Review Research},
  volume={7},
  number={2},
  pages={023038},
  year={2025},
  publisher={APS},
 doi = {https://doi.org/10.1103/PhysRevResearch.7.023038}
}

@article{bianchi2025nonlinear,
  title={Nonlinear protocol for high-dimensional quantum teleportation},
  author={Bianchi, Luca and Marconi, Carlo and Guarda, Giulia and Bacco, Davide},
  journal={Physical Review A},
  volume={112},
  number={1},
  pages={012615},
  year={2025},
  publisher={APS},
  doi = {https://doi.org/10.1103/3x44-664w}
}

@article{sephton2023quantum,
  title={Quantum transport of high-dimensional spatial information with a nonlinear detector},
  author={Sephton, Bereneice and Vall{\'e}s, Adam and Nape, Isaac and Cox, Mitchell A and Steinlechner, Fabian and Konrad, Thomas and Torres, Juan P and Roux, Filippus S and Forbes, Andrew},
  journal={Nature communications},
  volume={14},
  number={1},
  pages={8243},
  year={2023},
  publisher={Nature Publishing Group UK London},
doi={https://doi.org/10.1038/s41467-023-43949-x}
}

@article{nloswap,
  title={Faithful entanglement swapping based on sum-frequency generation},
  author={Sangouard, Nicolas and Sanguinetti, Bruno and Curtz, No{\'e} and Gisin, Nicolas and Thew, Rob and Zbinden, Hugo},
  journal={Physical review letters},
  volume={106},
  number={12},
  pages={120403},
  year={2011},
  publisher={APS},
  doi={10.1103/PhysRevLett.106.120403}
}

@article{zukowski1993event,
  title={‘‘Event-ready-detectors’’Bell experiment via entanglement swapping},
  author={{\.Z}ukowski, Marek and Zeilinger, Anton and Horne, Michael A and Ekert, Artur K},
  journal={Physical Review Letters},
  volume={71},
  number={26},
  pages={4287},
  year={1993},
  publisher={APS},
  doi={10.1103/PhysRevLett.71.4287}
}

@article{bartolucci2023fusion,
  title={Fusion-based quantum computation},
  author={Bartolucci, Sara and Birchall, Patrick and Bombin, Hector and Cable, Hugo and Dawson, Chris and Gimeno-Segovia, Mercedes and Johnston, Eric and Kieling, Konrad and Nickerson, Naomi and Pant, Mihir and others},
  journal={Nature Communications},
  volume={14},
  number={1},
  pages={912},
  year={2023},
  publisher={Nature Publishing Group UK London},
doi={https://doi.org/10.1038/s41467-023-36493-1}
}

@article{fusion1,
  title={Topological quantum computing with a very noisy network and local error rates approaching one percent},
  author={Nickerson, Naomi H and Li, Ying and Benjamin, Simon C},
  journal={Nature communications},
  volume={4},
  number={1},
  pages={1756},
  year={2013},
  publisher={Nature Publishing Group UK London},
    doi={https://doi.org/10.1038/ncomms2773}
}

@article{fusion2,
  title={From three-photon Greenberger-Horne-Zeilinger states to ballistic universal quantum computation},
  author={Gimeno-Segovia, Mercedes and Shadbolt, Pete and Browne, Dan E and Rudolph, Terry},
  journal={Physical review letters},
  volume={115},
  number={2},
  pages={020502},
  year={2015},
  publisher={APS},
  doi= {https://doi.org/10.1103/PhysRevLett.115.020502}
}

@article{fusion3,
  title={Very-large-scale integrated quantum graph photonics},
  author={Bao, Jueming and Fu, Zhaorong and Pramanik, Tanumoy and Mao, Jun and Chi, Yulin and Cao, Yingkang and Zhai, Chonghao and Mao, Yifei and Dai, Tianxiang and Chen, Xiaojiong and others},
  journal={Nature Photonics},
  volume={17},
  number={7},
  pages={573--581},
  year={2023},
  publisher={Nature Publishing Group UK London},
  doi={https://doi.org/10.1038/s41566-023-01187-z}
}

@article{stefano1,
  title={Scheme for universal high-dimensional quantum computation with linear optics},
  author={Paesani, Stefano and Bulmer, Jacob FF and Jones, Alex E and Santagati, Raffaele and Laing, Anthony},
  journal={Physical Review Letters},
  volume={126},
  number={23},
  pages={230504},
  year={2021},
  publisher={APS},
doi={10.1103/PhysRevLett.126.230504}
}

@article{stefano2,
  title={Tailoring fusion-based photonic quantum computing schemes to quantum emitters},
  author={Chan, Ming Lai and Bell, Thomas J and Pettersson, Love A and Chen, Susan X and Yard, Patrick and S{\o}rensen, Anders S and Paesani, Stefano},
  journal={PRX Quantum},
  volume={6},
  number={2},
  pages={020304},
  year={2025},
  publisher={APS},
  doi={10.1103/PRXQuantum.6.020304}
}

@article{khabiboulline2019optical,
  title={Optical interferometry with quantum networks},
  author={Khabiboulline, Emil T and Borregaard, Johannes and De Greve, Kristiaan and Lukin, Mikhail D},
  journal={Physical review letters},
  volume={123},
  number={7},
  pages={070504},
  year={2019},
  publisher={APS},
    doi={ https://doi.org/10.1103/PhysRevLett.123.070504}
}

@article{erhard2018twisted,
  title={Twisted photons: new quantum perspectives in high dimensions},
  author={Erhard, Manuel and Fickler, Robert and Krenn, Mario and Zeilinger, Anton},
  journal={Light: Science \& Applications},
  volume={7},
  number={3},
  pages={17146--17146},
  year={2018},
  publisher={Nature Publishing Group},
  doi={https://doi.org/10.1038/lsa.2017.146}
}

@article{xie2017quantum,
  title={The quantum Rabi model: solution and dynamics},
  author={Xie, Qiongtao and Zhong, Honghua and Batchelor, Murray T and Lee, Chaohong},
  journal={Journal of Physics A: Mathematical and Theoretical},
  volume={50},
  number={11},
  pages={113001},
  year={2017},
  publisher={IOP Publishing},
 doi= {10.1088/1751-8121/aa5a65}
}

@article{calsamiglia2001maximum,
  title={Maximum efficiency of a linear-optical Bell-state analyzer},
  author={Calsamiglia, John and L{\"u}tkenhaus, Norbert},
  journal={Applied Physics B},
  volume={72},
  pages={67--71},
  year={2001},
  publisher={Springer},
  doi={10.1007/s003400000484}
}

@article{vaidman1999methods,
  title={Methods for reliable teleportation},
  author={Vaidman, Lev and Yoran, Nadav},
  journal={Physical Review A},
  volume={59},
  number={1},
  pages={116},
  year={1999},
  publisher={APS},
  doi={10.1103/PhysRevA.59.116}
}

@article{paris2000optical,
  title={Optical Bell measurement by Fock filtering},
  author={Paris, MGA and Plenio, MB and Bose, S and Jonathan, D and D'Ariano, GM},
  journal={Physics Letters A},
  volume={273},
  number={3},
  pages={153--158},
  year={2000},
  publisher={Elsevier},
  doi={10.1016/S0375-9601(00)00477-1}
}

@book{furusawa2011quantum,
  title={Quantum teleportation and entanglement: a hybrid approach to optical quantum information processing},
  author={Furusawa, Akira and Van Loock, Peter},
  year={2011},
  publisher={John Wiley \& Sons}, 
  doi={10.1002/9783527635283}
}

@article{stefano4,
  title={High-threshold quantum computing by fusing one-dimensional cluster states},
  author={Paesani, Stefano and Brown, Benjamin J},
  journal={Physical Review Letters},
  volume={131},
  number={12},
  pages={120603},
  year={2023},
  publisher={APS},
  doi = {
https://doi.org/10.1103/PhysRevLett.131.120603
}
}

@article{purification,
  author    = {Charles H. Bennett and Gilles Brassard and Sandu Popescu and Benjamin Schumacher and John A. Smolin and William K. Wootters},
  title     = {Purification of Noisy Entanglement and Faithful Teleportation via Noisy Channels},
  journal   = {Physical Review Letters},
  volume    = {76},
  number    = {5},
  pages     = {722--725},
  year      = {1996},
  doi       = {10.1103/PhysRevLett.76.722}
}

@article{Distribution,
  author    = {Artur K. Ekert},
  title     = {Quantum Cryptography Based on Bell's Theorem},
  journal   = {Physical Review Letters},
  volume    = {67},
  number    = {6},
  pages     = {661--663},
  year      = {1991},
  doi       = {10.1103/PhysRevLett.67.661}
}

@article{Pan1998,
  author    = {Jian-Wei Pan et al.},
  title     = {Experimental Entanglement Swapping: Entangling Photons That Never Interacted},
  journal   = {Physical Review Letters},
  volume    = {80},
  pages     = {3891--3894},
  year      = {1998},
  doi       = {10.1103/PhysRevLett.80.3891}
}

@article{bouchard2018quantum,
  title={Quantum cryptography with twisted photons through an outdoor underwater channel},
  author={Bouchard, Fr{\'e}d{\'e}ric and Sit, Alicia and Hufnagel, Felix and Abbas, Aazad and Zhang, Yingwen and Heshami, Khabat and Fickler, Robert and Marquardt, Christoph and Leuchs, Gerd and Boyd, Robert w and others},
  journal={Optics express},
  volume={26},
  number={17},
  pages={22563--22573},
  year={2018},
  publisher={Optical Society of America},
  doi={10.1364/OE.26.022563}
}

@article{vagniluca2020efficient,
  title={Efficient time-bin encoding for practical high-dimensional quantum key distribution},
  author={Vagniluca, Ilaria and Da Lio, Beatrice and Rusca, Davide and Cozzolino, Daniele and Ding, Yunhong and Zbinden, Hugo and Zavatta, Alessandro and Oxenl{\o}we, Leif K and Bacco, Davide},
  journal={Physical Review Applied},
  volume={14},
  number={1},
  pages={014051},
  year={2020},
  publisher={APS},
 doi={10.1103/PhysRevApplied.14.014051}
}

@article{nadlinger2022experimental,
  title={Experimental quantum key distribution certified by Bell's theorem},
  author={Nadlinger, David P and Drmota, Peter and Nichol, Bethan C and Araneda, Gabriel and Main, Dougal and Srinivas, Raghavendra and Lucas, David M and Ballance, Christopher J and Ivanov, Kirill and Tan, EY-Z and others},
  journal={Nature},
  volume={607},
  number={7920},
  pages={682--686},
  year={2022},
  publisher={Nature Publishing Group UK London},
  doi={10.1038/s41586-022-04941-5}
}

@article{jo2019enhanced,
  title={Enhanced Bell state measurement for efficient measurement-device-independent quantum key distribution using 3-dimensional quantum states},
  author={Jo, Yonggi and Bae, Kwangil and Son, Wonmin},
  journal={Scientific reports},
  volume={9},
  number={1},
  pages={687},
  year={2019},
  publisher={Nature Publishing Group UK London},
  doi={10.1038/s41598-018-36513-x}
}

@article{lydersen2010hacking,
  title={Hacking commercial quantum cryptography systems by tailored bright illumination},
  author={Lydersen, Lars and Wiechers, Carlos and Wittmann, Christoffer and Elser, Dominique and Skaar, Johannes and Makarov, Vadim},
  journal={Nature photonics},
  volume={4},
  number={10},
  pages={686--689},
  year={2010},
  publisher={Nature Publishing Group},
  doi={10.1038/nphoton.2010.214}
}

@article{zapatero2023advances,
  title={Advances in device-independent quantum key distribution},
  author={Zapatero, V{\'\i}ctor and van Leent, Tim and Arnon-Friedman, Rotem and Liu, Wen-Zhao and Zhang, Qiang and Weinfurter, Harald and Curty, Marcos},
  journal={npj quantum information},
  volume={9},
  number={1},
  pages={10},
  year={2023},
  publisher={Nature Publishing Group UK London},
  doi={10.1038/s41534-023-00684-x}
}

@article{ren2014adaptive,
  title={Adaptive-optics-based simultaneous pre-and post-turbulence compensation of multiple orbital-angular-momentum beams in a bidirectional free-space optical link},
  author={Ren, Yongxiong and Xie, Guodong and Huang, Hao and Ahmed, Nisar and Yan, Yan and Li, Long and Bao, Changjing and Lavery, Martin PJ and Tur, Moshe and Neifeld, Mark A and others},
  journal={Optica},
  volume={1},
  number={6},
  pages={376--382},
  year={2014},
  publisher={Optical Society of America},
  doi={10.1364/OPTICA.1.000376}
}

@article{xu2013practical,
  title={Practical aspects of measurement-device-independent quantum key distribution},
  author={Xu, Feihu and Curty, Marcos and Qi, Bing and Lo, Hoi-Kwong},
  journal={New Journal of Physics},
  volume={15},
  number={11},
  pages={113007},
  year={2013},
  publisher={IOP Publishing},
  doi={10.1088/1367-2630/15/11/113007}
}

@article{liu2022toward,
  title={Toward a photonic demonstration of device-independent quantum key distribution},
  author={Liu, Wen-Zhao and Zhang, Yu-Zhe and Zhen, Yi-Zheng and Li, Ming-Han and Liu, Yang and Fan, Jingyun and Xu, Feihu and Zhang, Qiang and Pan, Jian-Wei},
  journal={Physical Review Letters},
  volume={129},
  number={5},
  pages={050502},
  year={2022},
  publisher={APS},
  doi={10.1103/PhysRevLett.129.050502}
}

@article{zhang2022device,
  title={A device-independent quantum key distribution system for distant users},
  author={Zhang, Wei and van Leent, Tim and Redeker, Kai and Garthoff, Robert and Schwonnek, Ren{\'e} and Fertig, Florian and Eppelt, Sebastian and Rosenfeld, Wenjamin and Scarani, Valerio and Lim, Charles C-W and others},
  journal={Nature},
  volume={607},
  number={7920},
  pages={687--691},
  year={2022},
  publisher={Nature Publishing Group UK London},
  doi={10.1038/s41586-022-04891-y}
}

@article{ding2017high,
  title={High-dimensional quantum key distribution based on multicore fiber using silicon photonic integrated circuits},
  author={Ding, Yunhong and Bacco, Davide and Dalgaard, Kjeld and Cai, Xinlun and Zhou, Xiaoqi and Rottwitt, Karsten and Oxenl{\o}we, Leif Katsuo},
  journal={npj Quantum Information},
  volume={3},
  number={1},
  pages={25},
  year={2017},
  publisher={Nature Publishing Group UK London},
    doi={
https://doi.org/10.1038/s41534-017-0026-2}
}

@article{mirhosseini2015,
  title={High-dimensional quantum cryptography with twisted light},
  author={Mirhosseini, Mohammad and Maga{\~n}a-Loaiza, Omar S and O’Sullivan, Malcolm N and Rodenburg, Brandon and Malik, Mehul and Lavery, Martin PJ and Padgett, Miles J and Gauthier, Daniel J and Boyd, Robert W},
  journal={New Journal of Physics},
  volume={17},
  number={3},
  pages={033033},
  year={2015},
  publisher={IOP Publishing},
 doi={10.1088/1367-2630/17/3/033033}
}

@article{vidal2000entanglement,
  title={Entanglement monotones},
  author={Vidal, Guifr{\'e}},
  journal={Journal of Modern Optics},
  volume={47},
  number={2-3},
  pages={355--376},
  year={2000},
  publisher={Taylor \& Francis},
  doi={10.1080/09500340008244048}
}

@article{acin2001classification,
  title={Classification of mixed three-qubit states},
  author={Ac{\'\i}n, Antonio and Bru{\ss}, Dagmar and Lewenstein, Maciej and Sanpera, Anna},
  journal={Physical Review Letters},
  volume={87},
  number={4},
  pages={040401},
  year={2001},
  publisher={APS},
  doi={10.1103/PhysRevLett.87.040401}
}

@article{dur2000three,
  title={Three qubits can be entangled in two inequivalent ways},
  author={D{\"u}r, Wolfgang and Vidal, Guifre and Cirac, J Ignacio},
  journal={Physical Review A},
  volume={62},
  number={6},
  pages={062314},
  year={2000},
  publisher={APS},
  doi={10.1103/PhysRevA.62.062314}
}

@article{bennett1992communication,
  title={Communication via one-and two-particle operators on Einstein-Podolsky-Rosen states},
  author={Bennett, Charles H and Wiesner, Stephen J},
  journal={Physical review letters},
  volume={69},
  number={20},
  pages={2881},
  year={1992},
  publisher={APS},
  doi={10.1103/PhysRevLett.69.2881}
}

@incollection{plenio2014introduction,
  title={An introduction to entanglement theory},
  author={Plenio, Martin B and Virmani, Shashank S},
  booktitle={Quantum information and coherence},
  pages={173--209},
  year={2014},
  publisher={Springer},
  doi={10.1007/978-3-319-04063-9_8}
}

@article{bennett2014quantum,
  title={Quantum cryptography: Public key distribution and coin tossing},
  author={Bennett, Charles H and Brassard, Gilles},
  journal={Theoretical computer science},
  volume={560},
  pages={7--11},
  year={2014},
  publisher={Elsevier},
  doi={10.1016/j.tcs.2014.05.025}
}

@article{mattle1996dense,
  title={Dense coding in experimental quantum communication},
  author={Mattle, Klaus and Weinfurter, Harald and Kwiat, Paul G and Zeilinger, Anton},
  journal={Physical review letters},
  volume={76},
  number={25},
  pages={4656},
  year={1996},
  publisher={APS},
  doi={10.1103/PhysRevLett.76.4656}
}

@article{hong1987measurement,
  title={Measurement of subpicosecond time intervals between two photons by interference},
  author={Hong, Chong-Ki and Ou, Zhe-Yu and Mandel, Leonard},
  journal={Physical review letters},
  volume={59},
  number={18},
  pages={2044},
  year={1987},
  publisher={APS},
 doi= {https://doi.org/10.1103/PhysRevLett.59.2044}
}

@article{carolan2015universal,
  title={Universal linear optics},
  author={Carolan, Jacques and Harrold, Christopher and Sparrow, Chris and Mart{\'\i}n-L{\'o}pez, Enrique and Russell, Nicholas J and Silverstone, Joshua W and Shadbolt, Peter J and Matsuda, Nobuyuki and Oguma, Manabu and Itoh, Mikitaka and others},
  journal={Science},
  volume={349},
  number={6249},
  pages={711--716},
  year={2015},
  publisher={American Association for the Advancement of Science},
doi={10.1126/science.aab3642}
}

@article{nolleke2013efficient,
  title={Efficient teleportation between remote single-atom quantum memories},
  author={N{\"o}lleke, Christian and Neuzner, Andreas and Reiserer, Andreas and Hahn, Carolin and Rempe, Gerhard and Ritter, Stephan},
  journal={Physical review letters},
  volume={110},
  number={14},
  pages={140403},
  year={2013},
  publisher={APS},
  doi={10.1103/PhysRevLett.110.140403}
}

@article{specht2011single,
  title={A single-atom quantum memory},
  author={Specht, Holger P and N{\"o}lleke, Christian and Reiserer, Andreas and Uphoff, Manuel and Figueroa, Eden and Ritter, Stephan and Rempe, Gerhard},
  journal={Nature},
  volume={473},
  number={7346},
  pages={190--193},
  year={2011},
  publisher={Nature Publishing Group UK London},
  doi={https://doi.org/10.1038/nature09997}
}

@article{walther2006cavity,
  title={Cavity quantum electrodynamics},
  author={Walther, Herbert and Varcoe, Benjamin TH and Englert, Berthold-Georg and Becker, Thomas},
  journal={Reports on Progress in Physics},
  volume={69},
  number={5},
  pages={1325},
  year={2006},
  publisher={IOP Publishing},
 doi={10.1088/0034-4885/69/5/R02}
}

@article{braunstein1998teleportation,
  title={Teleportation of continuous quantum variables},
  author={Braunstein, Samuel L and Kimble, H Jeff},
  journal={Physical review letters},
  volume={80},
  number={4},
  pages={869},
  year={1998},
  publisher={APS},
  doi={10.1103/PhysRevLett.80.869}
}

@article{furusawa1998unconditional,
  title={Unconditional quantum teleportation},
  author={Furusawa, Akira and S{\o}rensen, Jens Lykke and Braunstein, Samuel L and Fuchs, Christopher A and Kimble, H Jeff and Polzik, Eugene S},
  journal={science},
  volume={282},
  number={5389},
  pages={706--709},
  year={1998},
  publisher={American Association for the Advancement of Science},
  doi={10.1126/science.282.5389.706 }
}

@article{lloyd2001long,
  title={Long distance, unconditional teleportation of atomic states via complete Bell state measurements},
  author={Lloyd, S and Shahriar, MS and Shapiro, JH and Hemmer, PR},
  journal={Physical Review Letters},
  volume={87},
  number={16},
  pages={167903},
  year={2001},
  publisher={APS},
 doi = {https://doi.org/10.1103/PhysRevLett.87.167903}
}

@article{reiserer2015cavity,
  title={Cavity-based quantum networks with single atoms and optical photons},
  author={Reiserer, Andreas and Rempe, Gerhard},
  journal={Reviews of Modern Physics},
  volume={87},
  number={4},
  pages={1379--1418},
  year={2015},
  publisher={APS},
 doi={https://doi.org/10.1103/RevModPhys.87.1379}
}

@article{chen2024high,
  title={High-dimensional two-photon quantum controlled phase-flip gate},
  author={Chen, Mingyuan and Tang, Jiang-Shan and Cai, Miao and Lu, Yanqing and Nori, Franco and Xia, Keyu},
  journal={Physical Review Research},
  volume={6},
  number={3},
  pages={033004},
  year={2024},
  publisher={APS},
 doi={10.1103/PhysRevResearch.6.033004}
}

@article{xin2023qumode,
  title={Qumode fusion for continuous-variable cluster states},
  author={Xin, Jun},
  journal={Physical Review A},
  volume={108},
  number={2},
  pages={022406},
  year={2023},
  publisher={APS},
 doi={10.1103/PhysRevA.108.022406}
}

@article{duan2004scalable,
  title={Scalable photonic quantum computation through cavity-assisted interactions},
  author={Duan, L-M and Kimble, HJ},
  journal={Physical review letters},
  volume={92},
  number={12},
  pages={127902},
  year={2004},
  publisher={APS},
 doi={https://doi.org/10.1103/PhysRevLett.92.127902}
}

@article{welte2021nondestructive,
  title={A nondestructive Bell-state measurement on two distant atomic qubits},
  author={Welte, Stephan and Thomas, Philip and Hartung, Lukas and Daiss, Severin and Langenfeld, Stefan and Morin, Olivier and Rempe, Gerhard and Distante, Emanuele},
  journal={Nature Photonics},
  volume={15},
  number={7},
  pages={504--509},
  year={2021},
  publisher={Nature Publishing Group UK London},
 doi={https://doi.org/10.1038/s41566-021-00802-1}
}

@article{tang2014measurement,
  title={Measurement-device-independent quantum key distribution over 200 km},
  author={Tang, Yan-Lin and Yin, Hua-Lei and Chen, Si-Jing and Liu, Yang and Zhang, Wei-Jun and Jiang, Xiao and Zhang, Lu and Wang, Jian and You, Li-Xing and Guan, Jian-Yu and others},
  journal={Physical review letters},
  volume={113},
  number={19},
  pages={190501},
  year={2014},
  publisher={APS},
  doi={10.1103/PhysRevLett.113.190501}
}

@article{wang2021measurement,
  title={Measurement-device-independent quantum key distribution with leaky sources},
  author={Wang, Weilong and Tamaki, Kiyoshi and Curty, Marcos},
  journal={Scientific reports},
  volume={11},
  number={1},
  pages={1678},
  year={2021},
  publisher={Nature Publishing Group UK London},
  doi={10.1038/s41598-021-81003-2}
}

@article{aspelmeyer2003long,
  title={Long-distance quantum communication with entangled photons using satellites},
  author={Aspelmeyer, Markus and Jennewein, Thomas and Pfennigbauer, Martin and Leeb, Walter R and Zeilinger, Anton},
  journal={IEEE Journal of Selected Topics in Quantum Electronics},
  volume={9},
  number={6},
  pages={1541--1551},
  year={2003},
  publisher={IEEE},
  doi={10.1109/JSTQE.2003.820918}
}

@article{vallone2015experimental,
  title={Experimental satellite quantum communications},
  author={Vallone, Giuseppe and Bacco, Davide and Dequal, Daniele and Gaiarin, Simone and Luceri, Vincenza and Bianco, Giuseppe and Villoresi, Paolo},
  journal={Physical Review Letters},
  volume={115},
  number={4},
  pages={040502},
  year={2015},
  publisher={APS},
  doi={10.1103/PhysRevLett.115.040502}
}

@article{chen2022quantum,
  title={Quantum key distribution over 658 km fiber with distributed vibration sensing},
  author={Chen, Jiu-Peng and Zhang, Chi and Liu, Yang and Jiang, Cong and Zhao, Dong-Feng and Zhang, Wei-Jun and Chen, Fa-Xi and Li, Hao and You, Li-Xing and Wang, Zhen and others},
  journal={Physical Review Letters},
  volume={128},
  number={18},
  pages={180502},
  year={2022},
  publisher={APS},
  doi={10.1103/PhysRevLett.128.180502}
}

@article{wang2022twin,
  title={Twin-field quantum key distribution over 830-km fibre},
  author={Wang, Shuang and Yin, Zhen-Qiang and He, De-Yong and Chen, Wei and Wang, Rui-Qiang and Ye, Peng and Zhou, Yao and Fan-Yuan, Guan-Jie and Wang, Fang-Xiang and Chen, Wei and others},
  journal={Nature photonics},
  volume={16},
  number={2},
  pages={154--161},
  year={2022},
  publisher={Nature Publishing Group UK London},
  doi={10.1038/s41566-021-00928-2}
}

@article{yin2016measurement,
  title={Measurement-device-independent quantum key distribution over a 404 km optical fiber},
  author={Yin, Hua-Lei and Chen, Teng-Yun and Yu, Zong-Wen and Liu, Hui and You, Li-Xing and Zhou, Yi-Heng and Chen, Si-Jing and Mao, Yingqiu and Huang, Ming-Qi and Zhang, Wei-Jun and others},
  journal={Physical review letters},
  volume={117},
  number={19},
  pages={190501},
  year={2016},
  publisher={APS},
  doi={10.1103/PhysRevLett.117.190501}
}

@article{tang2016measurement,
  title={Measurement-device-independent quantum key distribution over untrustful metropolitan network},
  author={Tang, Yan-Lin and Yin, Hua-Lei and Zhao, Qi and Liu, Hui and Sun, Xiang-Xiang and Huang, Ming-Qi and Zhang, Wei-Jun and Chen, Si-Jing and Zhang, Lu and You, Li-Xing and others},
  journal={Physical Review X},
  volume={6},
  number={1},
  pages={011024},
  year={2016},
  publisher={APS},
  doi={10.1103/PhysRevX.6.011024}
}

@article{kamimaki2023deterministic,
  title={Deterministic Bell state measurement with a single quantum memory},
  author={Kamimaki, Akira and Wakamatsu, Keidai and Mikata, Kosuke and Sekiguchi, Yuhei and Kosaka, Hideo},
  journal={npj Quantum Information},
  volume={9},
  number={1},
  pages={101},
  year={2023},
  publisher={Nature Publishing Group UK London},
doi={https://doi.org/10.1038/s41534-023-00771-z}
}

@article{greentree2013fifty,
  title={Fifty years of Jaynes--Cummings physics},
  author={Greentree, Andrew D and Koch, Jens and Larson, Jonas},
  journal={Journal of Physics B: Atomic, Molecular and Optical Physics},
  volume={46},
  number={22},
  pages={220201},
  year={2013},
  publisher={IOP Publishing},
  doi={10.1088/0953-4075/46/22/220201}
}

@article{Lukens2017,
  author = {Lukens, Joseph M. and Lougovski, Pavel},
  title = {Frequency-encoded photonic qubits for scalable quantum information processing},
  journal = {Optica},
  volume = {4},
  pages = {8-16},
  year = {2017},
  doi={10.1364/OPTICA.4.000008}
}

@article{zhang2002quantum,
 title = {Quantum teleportation of light beams},
  author = {Zhang, T. C. and Goh, K. W. and Chou, C. W. and Lodahl, P. and Kimble, H. J.},
  journal = {Phys. Rev. A},
  volume = {67},
  issue = {3},
  pages = {033802},
  numpages = {16},
  year = {2003},
  month = {Mar},
  publisher = {American Physical Society},
  doi = {10.1103/PhysRevA.67.033802}
}

@article{takei2005high,
  title={High-Fidelity Teleportation beyond the No-Cloning Limit and Entanglement Swapping<? format?> for Continuous Variables},
  author={Takei, Nobuyuki and Yonezawa, Hidehiro and Aoki, Takao and Furusawa, Akira},
  journal={Physical review letters},
  volume={94},
  number={22},
  pages={220502},
  year={2005},
  publisher={APS},
  doi={10.1103/PhysRevLett.94.220502}
}

@article{jin2016simple,
  title={Simple method of generating and distributing frequency-entangled qudits},
  author={Jin, Rui-Bo and Shimizu, Ryosuke and Fujiwara, Mikio and Takeoka, Masahiro and Wakabayashi, Ryota and Yamashita, Taro and Miki, Shigehito and Terai, Hirotaka and Gerrits, Thomas and Sasaki, Masahide},
  journal={Quantum Science and Technology},
  volume={1},
  number={1},
  pages={015004},
  year={2016},
  publisher={IOP Publishing},
  doi={10.1088/2058-9565/1/1/015004}
}

@article{jo2019efficient,
  title={Efficient high-dimensional quantum key distribution with hybrid encoding},
  author={Jo, Yonggi and Park, Hee Su and Lee, Seung-Woo and Son, Wonmin},
  journal={Entropy},
  volume={21},
  number={1},
  pages={80},
  year={2019},
  publisher={MDPI},
 doi={https://doi.org/10.3390/e21010080}
}

@article{cheng2023high,
  title={High-dimensional time-frequency entanglement in a singly-filtered biphoton frequency comb},
  author={Cheng, Xiang and Chang, Kai-Chi and Sarihan, Murat Can and Mueller, Andrew and Spiropulu, Maria and Shaw, Matthew D and Korzh, Boris and Faraon, Andrei and Wong, Franco NC and Shapiro, Jeffrey H and others},
  journal={Communications Physics},
  volume={6},
  number={1},
  pages={278},
  year={2023},
  publisher={Nature Publishing Group UK London},
  doi={https://doi.org/10.1038/s42005-023-01370-2}
}

@article{stefano5,
  title={Deterministic generation of concatenated graph codes from quantum emitters},
  author={Pettersson, Love A and S{\o}rensen, Anders S and Paesani, Stefano},
  journal={PRX Quantum},
  volume={6},
  number={1},
  pages={010305},
  year={2025},
  publisher={APS},
  doi={https://doi.org/10.1103/PRXQuantum.6.010305}
}

@article{wootters1982single,
  title={A single quantum cannot be cloned},
  author={Wootters, William K and Zurek, Wojciech H},
  journal={Nature},
  volume={299},
  number={5886},
  pages={802--803},
  year={1982},
  publisher={Nature Publishing Group UK London},
  doi={10.1038/299802a0}
}

@article{briegel1998quantum,
  title={Quantum repeaters: the role of imperfect local operations in quantum communication},
  author={Briegel, Hans-J and D{\"u}r, Wolfgang and Cirac, Juan I and Zoller, Peter},
  journal={Physical Review Letters},
  volume={81},
  number={26},
  pages={5932},
  year={1998},
  publisher={APS},
  doi={10.1103/PhysRevLett.81.5932}
}

@article{sangouard2011quantum,
  title={Quantum repeaters based on atomic ensembles and linear optics},
  author={Sangouard, Nicolas and Simon, Christoph and De Riedmatten, Hugues and Gisin, Nicolas},
  journal={Reviews of Modern Physics},
  volume={83},
  number={1},
  pages={33},
  year={2011},
  publisher={APS},
  doi={10.1103/RevModPhys.83.33}
}

@article{muralidharan2014ultrafast,
  title={Ultrafast and fault-tolerant quantum communication across long distances},
  author={Muralidharan, Sreraman and Kim, Jungsang and L{\"u}tkenhaus, Norbert and Lukin, Mikhail D and Jiang, Liang},
  journal={Physical review letters},
  volume={112},
  number={25},
  pages={250501},
  year={2014},
  publisher={APS},
  doi={10.1103/PhysRevLett.112.250501}
}

@article{muralidharan2016optimal,
  title={Optimal architectures for long distance quantum communication},
  author={Muralidharan, Sreraman and Li, Linshu and Kim, Jungsang and L{\"u}tkenhaus, Norbert and Lukin, Mikhail D and Jiang, Liang},
  journal={Scientific reports},
  volume={6},
  number={1},
  pages={20463},
  year={2016},
  publisher={Nature Publishing Group UK London},
  doi={10.1038/srep20463}
}

@article{simon2010quantum,
  title={Quantum memories},
  author={Simon, Christoph and De Riedmatten, Hugues and Afzelius, Mikael and Sangouard, Nicolas and Zbinden, Hugo and Gisin, Nicolas},
  journal={The European Physical Journal D},
  volume={58},
  number={1},
  pages={1--22},
  year={2010},
  publisher={Springer},
  doi={10.1140/epjd/e2010-00103-y}
}

@article{barrett2005symmetry,
  title={Symmetry analyzer for nondestructive Bell-state detection using weak nonlinearities},
  author={Barrett, Sean D and Kok, Pieter and Nemoto, Kae and Beausoleil, Raymond G and Munro, William J and Spiller, Timothy P},
  journal={Physical Review A—Atomic, Molecular, and Optical Physics},
  volume={71},
  number={6},
  pages={060302},
  year={2005},
  publisher={APS},
  doi={10.1103/PhysRevA.71.060302}
}

@article{wang2016complete,
  title={Complete hyperentangled-Bell-state analysis for photonic qubits assisted by a three-level $\Lambda$-type system},
  author={Wang, Tie-Jun and Wang, Chuan},
  journal={Scientific reports},
  volume={6},
  number={1},
  pages={19497},
  year={2016},
  publisher={Nature Publishing Group UK London},
  doi={https://doi.org/10.1038/srep19497}
}

@article{sheng2010complete,
  title={Complete hyperentangled-Bell-state analysis for quantum communication},
  author={Sheng, Yu-Bo and Deng, Fu-Guo and Long, Gui Lu},
  journal={Physical Review A—Atomic, Molecular, and Optical Physics},
  volume={82},
  number={3},
  pages={032318},
  year={2010},
  publisher={APS},
  doi={10.1103/PhysRevA.82.032318}
}

@article{pisenti2011distinguishability,
  title={Distinguishability of hyperentangled Bell states by linear evolution and local projective measurement},
  author={Pisenti, N and Gaebler, CPE and Lynn, TW},
  journal={Physical Review A—Atomic, Molecular, and Optical Physics},
  volume={84},
  number={2},
  pages={022340},
  year={2011},
  publisher={APS},
  doi={10.1103/PhysRevA.84.022340}
}

@article{wei2007hyperentangled,
  title={Hyperentangled Bell-state analysis},
  author={Wei, Tzu-Chieh and Barreiro, Julio T and Kwiat, Paul G},
  journal={Physical Review A—Atomic, Molecular, and Optical Physics},
  volume={75},
  number={6},
  pages={060305},
  year={2007},
  publisher={APS},
  doi={10.1103/PhysRevA.75.060305}
}

@article{barbieri2007complete,
  title={Complete and deterministic discrimination of polarization Bell states assisted by momentum entanglement},
  author={Barbieri, Marco and Vallone, Giuseppe and Mataloni, Paolo and De Martini, Francesco},
  journal={Physical Review A—Atomic, Molecular, and Optical Physics},
  volume={75},
  number={4},
  pages={042317},
  year={2007},
  publisher={APS},
  doi={10.1103/PhysRevA.75.042317}
}

@article{schuck2006complete,
  title={Complete deterministic linear optics Bell state analysis},
  author={Schuck, Carsten and Huber, Gerhard and Kurtsiefer, Christian and Weinfurter, Harald},
  journal={Physical review letters},
  volume={96},
  number={19},
  pages={190501},
  year={2006},
  publisher={APS},
  doi={10.1103/PhysRevLett.96.190501}
}

@inproceedings{duc2024enhancement,
  title={Enhancement of teleportation average fidelity via photon addition operation},
  author={Duc, Truong Minh and Dat, Tran Quang},
  booktitle={Journal of Physics: Conference Series},
  volume={2744},
  pages={012002},
  year={2024},
  organization={IOP Publishing},
  doi={10.1088/1742-6596/2744/1/012002}
}

@article{kwiat1998embedded,
  title={Embedded Bell-state analysis},
  author={Kwiat, Paul G and Weinfurter, Harald},
  journal={Physical Review A},
  volume={58},
  number={4},
  pages={R2623},
  year={1998},
  publisher={APS},
  doi={10.1103/PhysRevA.58.R2623}
}

@article{azuma2015all,
  title={All-photonic quantum repeaters},
  author={Azuma, Koji and Tamaki, Kiyoshi and Lo, Hoi-Kwong},
  journal={Nature communications},
  volume={6},
  number={1},
  pages={6787},
  year={2015},
  publisher={Nature Publishing Group UK London},
  doi={10.1038/ncomms7787}
}

@article{shchukin2021waiting,
  title={Waiting time in quantum repeaters with probabilistic entanglement swapping},
  author={Shchukin, Evgeny and Schmidt, Ferdinand and van Loock, Peter},
  journal={Physical Review A},
  volume={100},
  number={3},
  pages={032322},
  year={2019},
  publisher={APS},
 doi={https://doi.org/10.1103/PhysRevA.100.032322}
}

@article{bacco2019high,
  title={High-dimensional quantum communication: Benefits, progress, and future challenges},
  author={Bacco, Davide and Da Lio, Beatrice and Cozzolino, Daniele and Oxenl{\o}we, Leif Katsuo},
  journal={Advanced Quantum Technologies},
  volume={2},
  number={12},
  pages={1900038},
  year={2019},
  publisher={Wiley Online Library},
  doi={10.1002/qute.201900038}
}

@article{gedik2015computational,
  title={Computational speed-up with a single qudit},
  author={Gedik, Zafer and Silva, Isabela A and {\c{C}}akmak, Bar{\i}s and Karpat, G{\"o}ktug and Vidoto, Edson Luiz G{\'e}a and Soares-Pinto, Diogo O and deAzevedo, Eduardo R and Fanchini, Felipe F},
  journal={Scientific reports},
  volume={5},
  number={1},
  pages={14671},
  year={2015},
  publisher={Nature Publishing Group UK London},
  doi = {https://doi.org/10.1038/srep14671}
}

@article{karacsony2024efficient,
  title={Efficient qudit based scheme for photonic quantum computing},
  author={Kar{\'a}csony, M{\'a}rton and Oroszl{\'a}ny, L{\'a}szl{\'o} and Zimboras, Zoltan},
  journal={SciPost Physics Core},
  volume={7},
  number={2},
  pages={032},
  year={2024},
  doi = {https://doi.org/10.48550/arXiv.2302.07357}
}

@article{luo2014geometry,
  title={Geometry of quantum computation with qudits},
  author={Luo, Ming-Xing and Chen, Xiu-Bo and Yang, Yi-Xian and Wang, Xiaojun},
  journal={Scientific reports},
  volume={4},
  number={1},
  pages={4044},
  year={2014},
  publisher={Nature Publishing Group UK London},
  doi={https://doi.org/10.1038/srep04044}
}

@article{wang2020qudits,
  title={Qudits and high-dimensional quantum computing},
  author={Wang, Yuchen and Hu, Zixuan and Sanders, Barry C and Kais, Sabre},
  journal={Frontiers in Physics},
  volume={8},
  pages={589504},
  year={2020},
  publisher={Frontiers Media SA},
    doi={
https://doi.org/10.3389/fphy.2020.589504}
}

@article{cesar2009extra,
  title={Extra phase noise from thermal fluctuations in nonlinear optical crystals},
  author={C{\'e}sar, J{\^o}natas Eduardo da Silva and Coelho, AS and Cassemiro, Katiuscia Nadyne and Villar, Alessandro de Sousa and Lassen, M and Nussenzveig, P and Martinelli, Marcelo},
  journal={Physical Review A—Atomic, Molecular, and Optical Physics},
  volume={79},
  number={6},
  pages={063816},
  year={2009},
  publisher={APS},
 doi={https://doi.org/10.1103/PhysRevA.79.063816}
}

@article{bacco2017space,
  title={Space division multiplexing chip-to-chip quantum key distribution},
  author={Bacco, Davide and Ding, Yunhong and Dalgaard, Kjeld and Rottwitt, Karsten and Oxenl{\o}we, Leif Katsuo},
  journal={Scientific reports},
  volume={7},
  number={1},
  pages={12459},
  year={2017},
  publisher={Nature Publishing Group UK London},
  doi={https://doi.org/10.1038/s41598-017-12309-3}
}

@article{sheridan2010security,
  title={Security proof for quantum key distribution using qudit systems},
  author={Sheridan, Lana and Scarani, Valerio},
  journal={Physical Review A—Atomic, Molecular, and Optical Physics},
  volume={82},
  number={3},
  pages={030301},
  year={2010},
  publisher={APS},
    doi={https://doi.org/10.1103/PhysRevA.82.030301}
}

@article{cerf2002security,
  title={Security of quantum key distribution using d-level systems},
  author={Cerf, Nicolas J and Bourennane, Mohamed and Karlsson, Anders and Gisin, Nicolas},
  journal={Physical review letters},
  volume={88},
  number={12},
  pages={127902},
  year={2002},
  publisher={APS},
    doi={https://doi.org/10.1103/PhysRevLett.88.127902}
}

@article{deng2017quantum,
  title={Quantum hyperentanglement and its applications in quantum information processing},
  author={Deng, Fu-Guo and Ren, Bao-Cang and Li, Xi-Han},
  journal={Science bulletin},
  volume={62},
  number={1},
  pages={46--68},
  year={2017},
  publisher={Elsevier},
  doi={10.1016/j.scib.2016.11.007}
}

@article{bechmann2000quantum,
  title={Quantum cryptography using larger alphabets},
  author={Bechmann-Pasquinucci, Helle and Tittel, Wolfgang},
  journal={Physical Review A},
  volume={61},
  number={6},
  pages={062308},
  year={2000},
  publisher={APS},
doi={https://doi.org/10.1103/PhysRevA.61.062308}
}

@article{ekert1991quantum,
  title={Quantum cryptography based on Bell’s theorem},
  author={Ekert, Artur K},
  journal={Physical review letters},
  volume={67},
  number={6},
  pages={661},
  year={1991},
  publisher={APS},
    doi={https://doi.org/10.1103/PhysRevLett.67.661}
}

@article{liu2009decay,
  title={Decay of multiqudit entanglement},
  author={Liu, Zhao and Fan, Heng},
  journal={Physical Review A—Atomic, Molecular, and Optical Physics},
  volume={79},
  number={6},
  pages={064305},
  year={2009},
  publisher={APS},
 doi={https://doi.org/10.1103/PhysRevA.79.064305}
}

@article{navez2003cloning,
  title={Cloning a real d-dimensional quantum state on the edge of the no-signaling condition},
  author={Navez, Patrick and Cerf, Nicolas J},
  journal={Physical Review A},
  volume={68},
  number={3},
  pages={032313},
  year={2003},
  publisher={APS},
  doi={https://doi.org/10.1103/PhysRevA.68.032313}
}

@article{bruss1999optimal,
  title={Optimal state estimation for d-dimensional quantum systems},
  author={Bru{\ss}, Dagmar and Macchiavello, Chiara},
  journal={Physics Letters A},
  volume={253},
  number={5-6},
  pages={249--251},
  year={1999},
  publisher={Elsevier},
  doi={
https://doi.org/10.1016/S0375-9601%2899%2900099-7}
}

@article{o2009photonic,
  title={Photonic quantum technologies},
  author={O'brien, Jeremy L and Furusawa, Akira and Vu{\v{c}}kovi{\'c}, Jelena},
  journal={Nature photonics},
  volume={3},
  number={12},
  pages={687--695},
  year={2009},
  publisher={Nature Publishing Group UK London},
  doi={https://doi.org/10.1038/nphoton.2009.229}
}

@article{yang2025programmable,
 title={Programmable quantum circuits in a large-scale photonic waveguide array},
  author={Yang, Yang and Chapman, Robert J and Youssry, Akram and Haylock, Ben and Lenzini, Francesco and Lobino, Mirko and Peruzzo, Alberto},
  journal={npj Quantum Information},
  volume={11},
  number={1},
  pages={19},
  year={2025},
  publisher={Nature Publishing Group UK London},
  doi={10.1038/s41534-024-00934-6}
}

@article{braunstein2005quantum,
  title={Quantum information with continuous variables},
  author={Braunstein, Samuel L and Van Loock, Peter},
  journal={Reviews of modern physics},
  volume={77},
  number={2},
  pages={513--577},
  year={2005},
  publisher={APS},
 doi ={https://doi.org/10.1103/RevModPhys.77.513}
}

@article{adesso2014continuous,
  title={Continuous variable quantum information: Gaussian states and beyond},
  author={Adesso, Gerardo and Ragy, Sammy and Lee, Antony R},
  journal={Open Systems \& Information Dynamics},
  volume={21},
  number={01n02},
  pages={1440001},
  year={2014},
  publisher={World Scientific},
 doi={https://doi.org/10.1142/S1230161214400010}
}

@article{andersen2015hybrid,
  title={Hybrid discrete-and continuous-variable quantum information},
  author={Andersen, Ulrik L and Neergaard-Nielsen, Jonas S and Van Loock, Peter and Furusawa, Akira},
  journal={Nature Physics},
  volume={11},
  number={9},
  pages={713--719},
  year={2015},
  publisher={Nature Publishing Group UK London},
 doi={https://doi.org/10.1038/nphys3410}
}

@book{serafini2023quantum,
  title={Quantum continuous variables: a primer of theoretical methods},
  author={Serafini, Alessio},
  year={2023},
  publisher={CRC press},
 doi={https://doi.org/10.1201/9781315118727}
}

@article{bowen2003experimental,
  title={Experimental investigation of continuous-variable quantum teleportation},
  author={Bowen, Warwick P and Treps, Nicolas and Buchler, Ben C and Schnabel, Roman and Ralph, Timothy C and Bachor, Hans-A and Symul, Thomas and Lam, Ping Koy},
  journal={Physical Review A},
  volume={67},
  number={3},
  pages={032302},
  year={2003},
  publisher={APS},
 doi={https://doi.org/10.1103/PhysRevA.67.032302}
}

@article{pirandola2006quantum,
  title={Quantum teleportation with continuous variables: A survey},
  author={Pirandola, Stefano and Mancini, Stefano},
  journal={Laser Physics},
  volume={16},
  pages={1418--1438},
  year={2006},
  publisher={Springer},
 doi={https://doi.org/10.1134/S1054660X06100057}
}

@book{vogel2006quantum,
  title={Quantum optics},
  author={Vogel, Werner and Welsch, Dirk-Gunnar},
  year={2006},
  publisher={John Wiley \& Sons},
 doi= {10.1002/3527608524}
}

@article{hofmann2000fidelity,
  title={Fidelity and information in the quantum teleportation of continuous variables},
  author={Hofmann, Holger F and Ide, Toshiki and Kobayashi, Takayoshi and Furusawa, Akira},
  journal={Physical Review A},
  volume={62},
  number={6},
  pages={062304},
  year={2000},
  publisher={APS},
 doi={https://doi.org/10.1103/PhysRevA.62.062304}
}

@article{lami2018gaussian,
  title={Gaussian quantum resource theories},
  author={Lami, Ludovico and Regula, Bartosz and Wang, Xin and Nichols, Rosanna and Winter, Andreas and Adesso, Gerardo},
  journal={Physical Review A},
  volume={98},
  number={2},
  pages={022335},
  year={2018},
  publisher={APS},
 doi= {https://doi.org/10.1103/PhysRevA.98.022335}
}

@article{bartlett2002efficient,
  title={Efficient classical simulation of continuous variable quantum information processes},
  author={Bartlett, Stephen D and Sanders, Barry C and Braunstein, Samuel L and Nemoto, Kae},
  journal={Physical Review Letters},
  volume={88},
  number={9},
  pages={097904},
  year={2002},
  publisher={APS},
 doi={https://doi.org/10.1103/PhysRevLett.88.097904}
}

@article{liu2024heralded,
  title={Heralded High-Dimensional Photon-Photon Quantum Gate},
  author={Liu, Zhi-Feng and Ren, Zhi-Cheng and Wan, Pei and Zhu, Wen-Zheng and Cheng, Zi-Mo and Wang, Jing and Shi, Yu-Peng and Xi, Han-Bing and Huber, Marcus and Friis, Nicolai and others},
  journal={arXiv preprint arXiv:2407.16356},
  year={2024},
doi =  {
https://doi.org/10.48550/arXiv.2407.16356}
}

@article{tang2025deterministic,
  title={Deterministic two-photon controlled-Z gate with the two-photon quantum Rabi model},
  author={Tang, Jia-Cheng and Zhao, Jin and Yang, Haitao and Tian, Junlong and Tang, Pinghua and Wang, Shuai-Peng and Lamata, Lucas and Peng, Jie},
  journal={Physical Review A},
  volume={111},
  number={5},
  pages={052601},
  year={2025},
  publisher={APS},
 doi={https://doi.org/10.1103/PhysRevA.111.052601}
}

@article{lloyd1999quantum,
  title={Quantum computation over continuous variables},
  author={Lloyd, Seth and Braunstein, Samuel L},
  journal={Physical Review Letters},
  volume={82},
  number={8},
  pages={1784},
  year={1999},
  publisher={APS},
 doi= {https://doi.org/10.1103/PhysRevLett.82.1784}
}

@article{liao2017satellite,
  title={Satellite-to-ground quantum key distribution},
  author={Liao, Sheng-Kai and Cai, Wen-Qi and Liu, Wei-Yue and Zhang, Liang and Li, Yang and Ren, Ji-Gang and Yin, Juan and Shen, Qi and Cao, Yuan and Li, Zheng-Ping and others},
  journal={Nature},
  volume={549},
  number={7670},
  pages={43--47},
  year={2017},
  publisher={Nature Publishing Group UK London},
  doi={10.1038/nature23655}
}

@inproceedings{mayers1998quantum,
  title={Quantum cryptography with imperfect apparatus},
  author={Mayers, Dominic and Yao, Andrew},
  booktitle={Proceedings 39th Annual Symposium on Foundations of Computer Science (Cat. No. 98CB36280)},
  pages={503--509},
  year={1998},
  organization={IEEE},
  doi={10.1109/SFCS.1998.743501}
}

@article{lo2012measurement,
  title={Measurement-device-independent quantum key distribution},
  author={Lo, Hoi-Kwong and Curty, Marcos and Qi, Bing},
  journal={Physical review letters},
  volume={108},
  number={13},
  pages={130503},
  year={2012},
  publisher={APS},
  doi={10.1038/s41598-021-81003-2}
}

@article{xu2020secure,
  title={Secure quantum key distribution with realistic devices},
  author={Xu, Feihu and Ma, Xiongfeng and Zhang, Qiang and Lo, Hoi-Kwong and Pan, Jian-Wei},
  journal={Reviews of modern physics},
  volume={92},
  number={2},
  pages={025002},
  year={2020},
  publisher={APS},
  doi={10.1103/RevModPhys.92.025002}
}

@article{pirandola2020advances,
  title={Advances in quantum cryptography},
  author={Pirandola, Stefano and Andersen, Ulrik L and Banchi, Leonardo and Berta, Mario and Bunandar, Darius and Colbeck, Roger and Englund, Dirk and Gehring, Tobias and Lupo, Cosmo and Ottaviani, Carlo and others},
  journal={Advances in optics and photonics},
  volume={12},
  number={4},
  pages={1012--1236},
  year={2020},
  publisher={Optical Society of America},
  doi={10.1364/AOP.361502}
}

@article{polkinghorne1999continuous,
  title={Continuous variable entanglement swapping},
  author={Polkinghorne, RES and Ralph, TC},
  journal={Physical review letters},
  volume={83},
  number={11},
  pages={2095},
  year={1999},
  publisher={APS},
 doi={https://doi.org/10.1103/PhysRevLett.83.2095}
}

@article{ide2001continuous,
  title={Continuous-variable teleportation of single-photon states},
  author={Ide, Toshiki and Hofmann, Holger F and Kobayashi, Takayoshi and Furusawa, Akira},
  journal={Physical Review A},
  volume={65},
  number={1},
  pages={012313},
  year={2001},
  publisher={APS},
 doi={https://doi.org/10.1103/PhysRevA.65.012313}
}

@article{hoelscher2011optimal,
  title={Optimal Gaussian entanglement swapping},
  author={Hoelscher-Obermaier, Jason and van Loock, Peter},
  journal={Physical Review A—Atomic, Molecular, and Optical Physics},
  volume={83},
  number={1},
  pages={012319},
  year={2011},
  publisher={APS},
 doi={https://doi.org/10.1103/PhysRevA.83.012319}
}

@article{hofer2013time,
  title={Time-continuous bell measurements},
  author={Hofer, Sebastian G and Vasilyev, Denis V and Aspelmeyer, Markus and Hammerer, Klemens},
  journal={Physical Review Letters},
  volume={111},
  number={17},
  pages={170404},
  year={2013},
  publisher={APS},
 doi={https://doi.org/10.1103/PhysRevLett.111.170404}
}

@article{dias2020quantum,
  title={Quantum repeater for continuous-variable entanglement distribution},
  author={Dias, Josephine and Winnel, Matthew S and Hosseinidehaj, Nedasadat and Ralph, Timothy C},
  journal={Physical Review A},
  volume={102},
  number={5},
  pages={052425},
  year={2020},
  publisher={APS},
 doi={https://doi.org/10.1103/PhysRevA.102.052425}
}

@article{liu2022all,
  title={All-optical entanglement swapping},
  author={Liu, Shengshuai and Lou, Yanbo and Chen, Yingxuan and Jing, Jietai},
  journal={Physical Review Letters},
  volume={128},
  number={6},
  pages={060503},
  year={2022},
  publisher={APS},
 doi={https://doi.org/10.1103/PhysRevLett.128.060503}
}

@article{menicucci2006universal,
  title={Universal quantum computation with continuous-variable cluster states},
  author={Menicucci, Nicolas C and Van Loock, Peter and Gu, Mile and Weedbrook, Christian and Ralph, <? format?> Timothy C and Nielsen, Michael A},
  journal={Physical review letters},
  volume={97},
  number={11},
  pages={110501},
  year={2006},
  publisher={APS}, 
  doi ={https://doi.org/10.1103/PhysRevLett.97.110501}
}

@article{gu2009quantum,
  title={Quantum computing with continuous-variable clusters},
  author={Gu, Mile and Weedbrook, Christian and Menicucci, Nicolas C and Ralph, Timothy C and van Loock, Peter},
  journal={Physical Review A—Atomic, Molecular, and Optical Physics},
  volume={79},
  number={6},
  pages={062318},
  year={2009},
  publisher={APS},
 doi= {https://doi.org/10.1103/PhysRevA.79.062318}
}

@article{menicucci2014fault,
  title={Fault-tolerant measurement-based quantum computing with continuous-variable cluster states},
  author={Menicucci, Nicolas C},
  journal={Physical review letters},
  volume={112},
  number={12},
  pages={120504},
  year={2014},
  publisher={APS},
 doi={https://doi.org/10.1103/PhysRevLett.112.120504}
}

@article{dur1999quantum,
  title={Quantum repeaters based on entanglement purification},
  author={D{\"u}r, Wolfgang and Briegel, H-J and Cirac, Juan Ignacio and Zoller, Peter},
  journal={Physical Review A},
  volume={59},
  number={1},
  pages={169},
  year={1999},
  publisher={APS},
 doi={https://doi.org/10.1103/PhysRevA.59.169}
}

@article{azuma2023quantum,
  title={Quantum repeaters: From quantum networks to the quantum internet},
  author={Azuma, Koji and Economou, Sophia E and Elkouss, David and Hilaire, Paul and Jiang, Liang and Lo, Hoi-Kwong and Tzitrin, Ilan},
  journal={Reviews of Modern Physics},
  volume={95},
  number={4},
  pages={045006},
  year={2023},
  publisher={APS},
 doi={https://doi.org/10.1103/RevModPhys.95.045006}
}

@article{pirandola2017fundamental,
  title={Fundamental limits of repeaterless quantum communications},
  author={Pirandola, Stefano and Laurenza, Riccardo and Ottaviani, Carlo and Banchi, Leonardo},
  journal={Nature communications},
  volume={8},
  number={1},
  pages={15043},
  year={2017},
  publisher={Nature Publishing Group UK London},
 doi ={https://doi.org/10.1038/ncomms15043}
}

@article{sangouard2009quantum,
  title={Quantum repeaters based on single trapped ions},
  author={Sangouard, Nicolas and Dubessy, Romain and Simon, Christoph},
  journal={Physical Review A—Atomic, Molecular, and Optical Physics},
  volume={79},
  number={4},
  pages={042340},
  year={2009},
  publisher={APS},
 doi={https://doi.org/10.1103/PhysRevA.79.042340}
}

@article{krutyanskiy2023telecom,
  title={Telecom-wavelength quantum repeater node based on a trapped-ion processor},
  author={Krutyanskiy, Victor and Canteri, Marco and Meraner, Martin and Bate, James and Krcmarsky, Vojtech and Schupp, Josef and Sangouard, Nicolas and Lanyon, Ben P},
  journal={Physical Review Letters},
  volume={130},
  number={21},
  pages={213601},
  year={2023},
  publisher={APS},
 doi={https://doi.org/10.1103/PhysRevLett.130.213601}
}

@article{chen2007fault,
  title={Fault-tolerant quantum repeater with atomic ensembles and linear optics},
  author={Chen, Zeng-Bing and Zhao, Bo and Chen, Yu-Ao and Schmiedmayer, J{\"o}rg and Pan, Jian-Wei},
  journal={Physical Review A—Atomic, Molecular, and Optical Physics},
  volume={76},
  number={2},
  pages={022329},
  year={2007},
  publisher={APS},
 doi={https://doi.org/10.1103/PhysRevA.76.022329}
}

@article{sangouard2008robust,
  title={Robust and efficient quantum repeaters with atomic ensembles and linear optics},
  author={Sangouard, Nicolas and Simon, Christoph and Zhao, Bo and Chen, Yu-Ao and De Riedmatten, Hugues and Pan, Jian-Wei and Gisin, Nicolas},
  journal={Physical Review A—Atomic, Molecular, and Optical Physics},
  volume={77},
  number={6},
  pages={062301},
  year={2008},
  publisher={APS},
  doi={https://doi.org/10.1103/PhysRevA.77.062301}
}

@article{han2010quantum,
  title={Quantum repeaters based on Rydberg-blockade-coupled atomic ensembles},
  author={Han, Yang and He, Bing and Heshami, Khabat and Li, Cheng-Zu and Simon, Christoph},
  journal={Physical Review A—Atomic, Molecular, and Optical Physics},
  volume={81},
  number={5},
  pages={052311},
  year={2010},
  publisher={APS},
  doi={https://doi.org/10.1103/PhysRevA.81.052311}
}

@article{jiang2009quantum,
  title={Quantum repeater with encoding},
  author={Jiang, Liang and Taylor, Jacob M and Nemoto, Kae and Munro, William J and Van Meter, Rodney and Lukin, Mikhail D},
  journal={Physical Review A—Atomic, Molecular, and Optical Physics},
  volume={79},
  number={3},
  pages={032325},
  year={2009},
  publisher={APS},
  doi= {https://doi.org/10.1103/PhysRevA.79.032325}
}

@article{van2006hybrid,
  title={Hybrid quantum repeater using bright coherent light},
  author={Van Loock, P and Ladd, TD and Sanaka, K and Yamaguchi, F and Nemoto, Kae and Munro, WJ and Yamamoto, Y},
  journal={Physical review letters},
  volume={96},
  number={24},
  pages={240501},
  year={2006},
  publisher={APS},
  doi={https://doi.org/10.1103/PhysRevLett.96.240501}
}

@article{ladd2006hybrid,
  title={Hybrid quantum repeater based on dispersive CQED interactions between matter qubits and bright coherent light},
  author={Ladd, Thaddeus D and van Loock, Peter and Nemoto, Kae and Munro, William J and Yamamoto, Yoshihisa},
  journal={New Journal of Physics},
  volume={8},
  number={9},
  pages={184},
  year={2006},
  publisher={IOP Publishing},
  doi= {10.1088/1367-2630/8/9/184}
}

@article{bergmann2019hybrid,
  title={Hybrid quantum repeater for qudits},
  author={Bergmann, Marcel and van Loock, Peter},
  journal={Physical Review A},
  volume={99},
  number={3},
  pages={032349},
  year={2019},
  publisher={APS},
  doi={https://doi.org/10.1103/PhysRevA.99.032349}
}

@article{li2019experimental,
  title={Experimental quantum repeater without quantum memory},
  author={Li, Zheng-Da and Zhang, Rui and Yin, Xu-Fei and Liu, Li-Zheng and Hu, Yi and Fang, Yu-Qiang and Fei, Yue-Yang and Jiang, Xiao and Zhang, Jun and Li, Li and others},
  journal={Nature photonics},
  volume={13},
  number={9},
  pages={644--648},
  year={2019},
  publisher={Nature Publishing Group UK London},
  doi= {https://doi.org/10.1038/s41566-019-0468-5}
}

@article{psiquantum2025manufacturable,
  title={A manufacturable platform for photonic quantum computing},
    author = {PsiQuantum team},
  journal={Nature},
  pages={1--3},
  year={2025},
  publisher={Nature Publishing Group UK London},
  doi={https://doi.org/10.1038/s41586-025-08820-7}
}

@article{song2024encoded,
  title={Encoded-Fusion-Based Quantum Computation for High Thresholds with Linear Optics},
  author={Song, Wooyeong and Kang, Nuri and Kim, Yong-Su and Lee, Seung-Woo},
  journal={Physical Review Letters},
  volume={133},
  number={5},
  pages={050605},
  year={2024},
  publisher={APS},
 doi={10.1103/PhysRevLett.133.050605}
}

@article{cocchi2025time,
  title={Time-bin encoding quantum key distribution in free-space horizontal links during nighttime and daytime},
  author={Cocchi, Sebastiano and Ribezzo, Domenico and Guarda, Giulia and Centorrino, Pietro and Occhipinti, Tommaso and Zavatta, Alessandro and Bacco, Davide},
  journal={Optica Quantum},
  volume={3},
  number={4},
  pages={346--350},
  year={2025},
  publisher={Optica Publishing Group},
 doi = {https://doi.org/10.1364/OPTICAQ.553977}
}

@article{mueller2025performance,
  title={Performance of Cascade and LDPC Codes for Information Reconciliation on Industrial Quantum Key Distribution Systems},
  author={Mueller, Ronny and De Lazzari, Claudia and Chirici, Fernando and Vagniluca, Ilaria and Oxenl{\o}we, Leif Katsuo and Forchhammer, S{\o}ren and Zavatta, Alessandro and Bacco, Davide},
  journal={IET Quantum Communication},
  volume={6},
  number={1},
  pages={e70003},
  year={2025},
  publisher={Wiley Online Library},
 doi = { https://doi.org/10.1049/qtc2.70003}
}

@article{ribezzo2023deploying,
  title={Deploying an inter-European quantum network},
  author={Ribezzo, Domenico and Zahidy, Mujtaba and Vagniluca, Ilaria and Biagi, Nicola and Francesconi, Saverio and Occhipinti, Tommaso and Oxenl{\o}we, Leif K and Lon{\v{c}}ari{\'c}, Martin and Cviti{\'c}, Ivan and Stip{\v{c}}evi{\'c}, Mario and others},
  journal={Advanced Quantum Technologies},
  volume={6},
  number={2},
  pages={2200061},
  year={2023},
  publisher={Wiley Online Library},
 doi = {https://doi.org/10.1002/qute.202200061}
}

@article{raussendorf2001one,
  title={A one-way quantum computer},
  author={Raussendorf, Robert and Briegel, Hans J},
  journal={Physical review letters},
  volume={86},
  number={22},
  pages={5188},
  year={2001},
  publisher={APS},
 doi = {https://doi.org/10.1103/PhysRevLett.86.5188}
}

@article{fowler2010surface,
  title={Surface code quantum communication},
  author={Fowler, Austin G and Wang, David S and Hill, Charles D and Ladd, Thaddeus D and Van Meter, <? format?> Rodney and Hollenberg, Lloyd CL},
  journal={Physical review letters},
  volume={104},
  number={18},
  pages={180503},
  year={2010},
  publisher={APS},
 doi = {https://doi.org/10.1103/PhysRevLett.104.180503}
}

@article{pant2017rate,
  title={Rate-distance tradeoff and resource costs for all-optical quantum repeaters},
  author={Pant, Mihir and Krovi, Hari and Englund, Dirk and Guha, Saikat},
  journal={Physical Review A},
  volume={95},
  number={1},
  pages={012304},
  year={2017},
  publisher={APS},
 doi = {https://doi.org/10.1103/PhysRevA.95.012304}
}

@article{ewert2016ultrafast,
  title={Ultrafast long-distance quantum communication with static linear optics},
  author={Ewert, Fabian and Bergmann, Marcel and Van Loock, Peter},
  journal={Physical review letters},
  volume={117},
  number={21},
  pages={210501},
  year={2016},
  publisher={APS}, 
 doi = {https://doi.org/10.1103/PhysRevLett.117.210501}
}

@article{lee2019fundamental,
  title={Fundamental building block for all-optical scalable quantum networks},
  author={Lee, Seung-Woo and Ralph, Timothy C and Jeong, Hyunseok},
  journal={Physical Review A},
  volume={100},
  number={5},
  pages={052303},
  year={2019},
  publisher={APS},
 doi = {https://doi.org/10.1103/PhysRevA.100.052303}
}

@article{zwerger2016measurement,
  title={Measurement-based quantum communication},
  author={Zwerger, M and Briegel, HJ and D{\"u}r, W},
  journal={Applied Physics B},
  volume={122},
  number={3},
  pages={50},
  year={2016},
  publisher={Springer},
 doi = {10.1007/s00340-015-6285-8}
}

@article{fukui2021all,
  title={All-optical long-distance quantum communication with Gottesman-Kitaev-Preskill qubits},
  author={Fukui, Kosuke and Alexander, Rafael N and van Loock, Peter},
  journal={Physical Review Research},
  volume={3},
  number={3},
  pages={033118},
  year={2021},
  publisher={APS},
doi = {https://doi.org/10.1103/PhysRevResearch.3.033118}
}

@misc{rozpedek2021quantum,
  title={Quantum repeaters based on concatenated bosonic and discrete-variable quantum codes},
  author={Rozpedek, Filip and Noh, Kyungjoo and Xu, Qian and Guha, Saikat and Jiang, Liang},
  year={2021},
  publisher={Jun},
 doi = {https://doi.org/10.1038/s41534-021-00438-7}
}

@article{chen2020parametric,
  title={Parametric amplifier for Bell measurement in continuous-variable quantum state teleportation},
  author={Chen, Xin and Ou, ZY},
  journal={Physical Review A},
  volume={102},
  number={3},
  pages={032407},
  year={2020},
  publisher={APS},
 doi = {https://doi.org/10.1103/PhysRevA.102.032407}
}

@article{du2025complete,
  title={A complete continuous-variable quantum computation architecture based on the 2D spatiotemporal cluster state},
  author={Du, Peilin and Zhang, Jing and Zhang, Tiancai and Yang, Rongguo and Gao, Jiangrui},
  journal={Scientific Reports},
  volume={15},
  number={1},
  pages={18199},
  year={2025},
  publisher={Nature Publishing Group UK London},
 doi = {https://doi.org/10.1038/s41598-025-02899-8}
}

@article{booth2023flow,
  title={Flow conditions for continuous variable measurement-based quantum computing},
  author={Booth, Robert I and Markham, Damian},
  journal={Quantum},
  volume={7},
  pages={1146},
  year={2023},
  publisher={Verein zur F{\"o}rderung des Open Access Publizierens in den Quantenwissenschaften},
 doi = {https://doi.org/10.22331/q-2023-10-19-1146}
}

@article{ustun2025fusion,
  title={Fusion for high-dimensional linear-optical quantum computing with improved success probability},
  author={{\"U}st{\"u}n, G{\"o}zde and Rieffel, Eleanor G and Devitt, Simon J and Saied, Jason},
  journal={Physical Review Applied},
  volume={24},
  number={4},
  pages={044024},
  year={2025},
  publisher={APS},
 doi = {https://doi.org/10.1103/l7bg-hc8c}
}

@article{zhou2015complete,
  title={Complete logic Bell-state analysis assisted with photonic Faraday rotation},
  author={Zhou, Lan and Sheng, Yu-Bo},
  journal={Physical Review A},
  volume={92},
  number={4},
  pages={042314},
  year={2015},
  publisher={APS},
 doi = {https://doi.org/10.1103/PhysRevA.92.042314}
}

@article{chang2025recent,
  title={Recent advances in high-dimensional quantum frequency combs},
  author={Chang, Kai-Chi and Cheng, Xiang and Sarihan, Murat Can and Wong, Chee Wei},
  journal={Newton},
  volume={1},
  number={1},
  year={2025},
  publisher={Elsevier},
 doi = {https://doi.org/10.1016/j.newton.2025.100024}
}

@article{chang2024time,
  title={Time-reversible and fully time-resolved ultra-narrowband biphoton frequency combs},
  author={Chang, Kai-Chi and Cheng, Xiang and Sarihan, Murat Can and Wong, Chee Wei},
  journal={APL Quantum},
  volume={1},
  number={1},
  year={2024},
  publisher={AIP Publishing},
 doi = {https://doi.org/10.1063/5.0180543}
}

@article{chang2023towards,
  title={Towards optimum Franson interference recurrence in mode-locked singly-filtered biphoton frequency combs},
  author={Chang, Kai-Chi and Cheng, Xiang and Sarihan, Murat Can and Wong, Chee Wei},
  journal={Photonics Research},
  volume={11},
  number={7},
  pages={1175--1184},
  year={2023},
  publisher={Chinese Laser Press and Optica Publishing Group},
 doi = {https://doi.org/10.1364/PRJ.483570
}
}

@article{chang2021648,
  title={648 Hilbert-space dimensionality in a biphoton frequency comb: entanglement of formation and Schmidt mode decomposition},
  author={Chang, Kai-Chi and Cheng, Xiang and Sarihan, Murat Can and Vinod, Abhinav Kumar and Lee, Yoo Seung and Zhong, Tian and Gong, Yan-Xiao and Xie, Zhenda and Shapiro, Jeffrey H and Wong, Franco NC and others},
  journal={npj Quantum Information},
  volume={7},
  number={1},
  pages={48},
  year={2021},
  publisher={Nature Publishing Group UK London},
 doi = {https://doi.org/10.1038/s41534-021-00388-0}
}

@article{xu2015discrete,
  title={Discrete and continuous variables for measurement-device-independent quantum cryptography},
  author={Xu, Feihu and Curty, Marcos and Qi, Bing and Qian, Li and Lo, Hoi-Kwong},
  journal={Nature Photonics},
  volume={9},
  number={12},
  pages={772--773},
  year={2015},
  publisher={Nature Publishing Group UK London},
 doi = {https://doi.org/10.1038/nphoton.2015.206}
}

@article{wang2012photonic,
  title={Photonic two-qubit parity gate with tiny cross--Kerr nonlinearity},
  author={Wang, Xin-Wen and Zhang, Deng-Yu and Tang, Shi-Qing and Xie, Li-Jun and Wang, Zhi-Yong and Kuang, Le-Man},
  journal={Physical Review A—Atomic, Molecular, and Optical Physics},
  volume={85},
  number={5},
  pages={052326},
  year={2012},
  publisher={APS},
doi = {https://doi.org/10.1103/PhysRevA.85.052326}
}

@article{lin2009quantum,
  title={Quantum control gates with weak cross-Kerr nonlinearity},
  author={Lin, Qing and Li, Jian},
  journal={Physical Review A—Atomic, Molecular, and Optical Physics},
  volume={79},
  number={2},
  pages={022301},
  year={2009},
  publisher={APS},
   doi = { https://doi.org/10.1103/PhysRevA.79.022301}
}

@article{vitali2000complete,
  title={Complete quantum teleportation with a Kerr nonlinearity},
  author={Vitali, David and Fortunato, Mauro and Tombesi, Paolo},
  journal={Physical review letters},
  volume={85},
  number={2},
  pages={445},
  year={2000},
  publisher={APS},
   doi = {https://doi.org/10.1103/PhysRevLett.85.445}
}

@article{kwiat1997hyper,
  title={Hyper-entangled states},
  author={Kwiat, Paul G},
  journal={Journal of modern optics},
  volume={44},
  number={11-12},
  pages={2173--2184},
  year={1997},
  publisher={Taylor \& Francis},
  doi = {https://doi.org/10.1080/09500349708231877}
}

@article{wang2015quantum,
  title={Quantum teleportation of multiple degrees of freedom of a single photon},
  author={Wang, Xi-Lin and Cai, Xin-Dong and Su, Zu-En and Chen, Ming-Cheng and Wu, Dian and Li, Li and Liu, Nai-Le and Lu, Chao-Yang and Pan, Jian-Wei},
  journal={Nature},
  volume={518},
  number={7540},
  pages={516--519},
  year={2015},
  publisher={Nature Publishing Group UK London},
   doi = {https://doi.org/10.1038/nature14246}
}

@article{bacco2021proposal,
  title={Proposal for practical multidimensional quantum networks},
  author={Bacco, Davide and Bulmer, Jacob FF and Erhard, Manuel and Huber, Marcus and Paesani, Stefano},
  journal={Physical Review A},
  volume={104},
  number={5},
  pages={052618},
  year={2021},
  publisher={APS},
   doi = {https://doi.org/10.1103/PhysRevA.104.052618}
}

@article{asenbeck2024hybrid,
  title={Hybrid approach to mitigate errors in linear photonic bell-state measurement for quantum interconnects},
  author={Asenbeck, Beate E and Kawasaki, Akito and Boyer, Ambroise and Darras, Tom and Urvoy, Alban and Furusawa, Akira and Laurat, Julien},
  journal={PRX Quantum},
  volume={5},
  number={3},
  pages={030331},
  year={2024},
  publisher={APS},
  doi= {https://doi.org/10.1103/PRXQuantum.5.030331}
}

@article{van2006experimental,
  title={Experimental quantum teleportation with a three-Bell-state analyzer},
  author={Van Houwelingen, JAW and Beveratos, Alexios and Brunner, Nicolas and Gisin, Nicolas and Zbinden, Hugo},
  journal={Physical Review A—Atomic, Molecular, and Optical Physics},
  volume={74},
  number={2},
  pages={022303},
  year={2006},
  publisher={APS},
  doi = {https://doi.org/10.1103/PhysRevA.74.022303}
}

@article{wu2025integration,
  title={Integration of quantum key distribution and high-throughput classical communications in field-deployed multi-core fibers},
  author={Wu, Qi and Ribezzo, Domenico and Di Sciullo, Giammarco and Cocchi, Sebastiano and Ann Shaji, Divya and Alves Zischler, Lucas and Luis, Ruben and Serena, Paolo and Lasagni, Chiara and Bononi, Alberto and others},
  journal={Light: Science \& Applications},
  volume={14},
  number={1},
  pages={274},
  year={2025},
  publisher={Nature Publishing Group UK London},
  doi = {https://doi.org/10.1038/s41377-025-01982-z}
}

@article{tomm2021bright,
  title={A bright and fast source of coherent single photons},
  author={Tomm, Natasha and Javadi, Alisa and Antoniadis, Nadia Olympia and Najer, Daniel and L{\"o}bl, Matthias Christian and Korsch, Alexander Rolf and Schott, R{\"u}diger and Valentin, Sascha Ren{\'e} and Wieck, Andreas Dirk and Ludwig, Arne and others},
  journal={Nature Nanotechnology},
  volume={16},
  number={4},
  pages={399--403},
  year={2021},
  publisher={Nature Publishing Group UK London},
  doi = {https://doi.org/10.1038/s41565-020-00831-x}
}

@article{zhai2022quantum,
  title={Quantum interference of identical photons from remote GaAs quantum dots},
  author={Zhai, Liang and Nguyen, Giang N and Spinnler, Clemens and Ritzmann, Julian and L{\"o}bl, Matthias C and Wieck, Andreas D and Ludwig, Arne and Javadi, Alisa and Warburton, Richard J},
  journal={Nature nanotechnology},
  volume={17},
  number={8},
  pages={829--833},
  year={2022},
  publisher={Nature Publishing Group UK London},
 doi = {https://doi.org/10.1038/s41565-022-01131-2}
}

@article{ding2025high,
  title={High-efficiency single-photon source above the loss-tolerant threshold for efficient linear optical quantum computing},
  author={Ding, Xing and Guo, Yong-Peng and Xu, Mo-Chi and Liu, Run-Ze and Zou, Geng-Yan and Zhao, Jun-Yi and Ge, Zhen-Xuan and Zhang, Qi-Hang and Liu, Hua-Liang and Wang, Lin-Jun and others},
  journal={Nature Photonics},
  volume={19},
  number={4},
  pages={387--391},
  year={2025},
  publisher={Nature Publishing Group UK London},
  doi ={https://doi.org/10.1038/s41566-025-01639-8}
}

@article{you2020superconducting,
  title={Superconducting nanowire single-photon detectors for quantum information},
  author={You, Lixing},
  journal={Nanophotonics},
  volume={9},
  number={9},
  pages={2673--2692},
  year={2020},
  publisher={De Gruyter},
  doi = { https://doi.org/10.1515/nanoph-2020-0186}
}

@article{venza2025research,
  title={Research trends in single-photon detectors based on superconducting wires},
  author={Venza, Francesco P and Colangelo, Marco},
  journal={APL Photonics},
  volume={10},
  number={4},
  year={2025},
  publisher={AIP Publishing},
  doi= {https://doi.org/10.1063/5.0246490}
}

@article{kong2024large,
  title={Large-inductance superconducting microstrip photon detector enabling 10 photon-number resolution},
  author={Kong, Ling-Dong and Zhang, Tian-Zhu and Liu, Xiao-Yu and Li, Hao and Wang, Zhen and Xie, Xiao-Ming and You, Li-Xing},
  journal={Advanced Photonics},
  volume={6},
  number={1},
  pages={016004--016004},
  year={2024},
  publisher={Society of Photo-Optical Instrumentation Engineers},
  doi = {https://doi.org/10.1117/1.AP.6.1.016004}
}

@article{ding2025photon,
  title={Photon-number-resolving single-photon detector with a system detection efficiency of 98\% and photon-number resolution of 32},
  author={Ding, Chaomeng and Zhang, Xingyu and Xiong, Jiamin and Xiao, You and Zhang, Tianzhu and Huang, Jia and Xu, Hongxin and Liu, Xiaoyu and You, Lixing and Wang, Zhen and others},
  journal={ACS Photonics},
  volume={12},
  number={9},
  pages={4924--4931},
  year={2025},
  publisher={ACS Publications},
  doi = {10.1021/acsphotonics.5c00508}
}
\end{document}